# Predicting room-temperature conductivity of Na-ion super ionic conductors with the minimal number of easily-accessible descriptors

*Seong-Hoon Jang,[1,2,*] Randy Jalem,[2] and Yoshitaka Tateyama[2,3]*

[1] Institute for Materials Research, Tohoku University, 2-1-1 Katahira, Aoba-ku, Sendai, 980-8577, Japan

[2] Research Center for Energy and Environmental Materials (GREEN), National Institute for Materials Science (NIMS), 1-1 Namiki, Tsukuba, Ibaraki 305-0044, Japan

[3] Laboratory for Chemistry and Life Science, Tokyo Institute of Technology, 4259 Nagatsuta, Midori-ku, Yokohama, 226-8501, Japan

*Corresponding author: jang.seonghoon.b4@tohoku.ac.jp

ABSTRACT. Given the vast compositional possibilities $Na_n M_m M'_{m'} Si_{3-p-a} P_p As_a O_{12}$, Na-ion superionic conductors (NASICON) are attractive but complicate for designing materials with enhanced room-temperature Na-ion conductivity $\sigma_{\text{Na,300K}}$. We propose an explicit regression model for $\sigma_{\text{Na,300K}}$ with easily-accessible descriptors, by exploiting density functional theory molecular dynamics (DFT-MD). Initially, we demonstrate that two primary descriptors, the bottleneck width along Na-ion diffusion paths $d_1$ and the average Na-Na distance $\langle d_{\text{Na}-\text{Na}} \rangle$, modulate room-temperature Na-ion self-diffusion coefficient $D_{\text{Na,300K}}$. Then, we introduce two secondary easily-accessible descriptors: Na-ion content $n$, which influences $d_1$, $\langle d_{\text{Na}-\text{Na}} \rangle$, and Na-ion density $\rho_{\text{Na}}$; and the average ionic radius $\langle r_M \rangle$ of metal ions, which impacts $d_1$ and $\langle d_{\text{Na}-\text{Na}} \rangle$. These secondary descriptors enable the development of a regression model for $\sigma_{\text{Na,300K}}$ with $n$ and $\langle r_M \rangle$ only. Subsequently, this model identifies a promising yet unexplored stable composition, $Na_{2.75}Zr_{1.75}Nb_{0.25}Si_2PO_{12}$, which, upon DFT-MD calculations, indeed exhibits $\sigma_{\text{Na,300K}} > 10^{-3}$ S·cm$^{-1}$. Furthermore, the adjusted version effectively fits 140 experimental values with $R^2 = 0.718$.

Inorganic solid electrolytes hold significant promise in advancing solid-state battery technology due to their possible superior electrochemical and thermal stability, as well as decreased flammability when compared to liquid electrolytes, if ideal.[1-5] Several classes of inorganic solid electrolytes have shown great potential, including Li-ion oxides [such as lithium superionic conductor (LISICON) $Li_{14}ZnGe_4O_{16}$[6] and LLZO $Li_7La_3Zr_2O_{12}$],[7] Li-ion sulfides (LGPS $Li_{10}GeP_2S_{12}$),[8] and Na-ion sulfides ($Na_nM_mM'_{m'}S_4$; $M$ and $M'$ are metal ions).[9, 10] Among these, Na-ion oxides known as Na-ion superionic conductors (NASICON) has attracted considerable attention due to the earth-abundant sodium, structural and thermal stability, and notably high ion conductivity facilitated by its unique three-dimensional "skeleton" structure.[11-13]

NASICON, represented by the generalized chemical formula $Na_nM_mM'_{m'}Si_{3-p-a}P_pAs_aO_{12}$, offers the ability to incorporate a wide range of metal ions ($M$ or $M'$) with varying valences and ionic radii, making it a compelling system for designing materials with enhanced Na-ionic conduction.[14] Numerous substitutions have been explored, including divalent ions such as Mg,[15-19] Ca,[19, 20] Fe,[21] Co,[17] Zn,[15, 18] Sr,[19] Cd,[18] and Ba;[19] trivalent ions such as Al,[18, 22-25] Sc,[23, 24, 26-30] Cr,[18, 23, 24, 26, 31-33] Fe,[18, 21, 23, 24, 26, 32, 33] Ga,[23, 24] Y,[23, 24, 34-37] In,[18, 23, 24, 31, 35, 38] Nd,[39] Gd,[35, 40, 41] Tb,[35] Dy,[35, 42] Er,[35, 42] and Yb;[18, 23, 24, 31, 35, 41-43] tetravalent ions such as Ti,[15, 25, 44-46] Ge,[25, 44, 45] Zr,[25, 45] Sn,[15, 25, 45, 47] Hf,[25, 44, 45, 48] Ce,[41, 44] and Th;[44] and pentavalent ions such as V,[15] Nb,[15, 28, 49-53] Mo,[50] and Ta.[15]

Given the extensive array of compositions within the framework of $Na_nM_mM'_{m'}Si_{3-p-a}P_pAs_aO_{12}$, we endeavor to construct an "explicit" regression model for the prediction of room-temperature Na-ion conductivity $\sigma_{Na,300K}$, utilizing two descriptors easily accessible from the chemical formula only, which is an unexplored avenue in the current literature.

Undertaking exhaustive experimental investigations for all conceivable compositions poses a formidable challenge; hence, we advocate for a theoretical approach, specifically employing density functional theory molecular dynamics (DFT-MD) in conjunction with regression modeling techniques. Herein, we extensively employ the beta regression modeling technique, which offers a flexible framework for analyzing data spanning a wide range of magnitudes in $\sigma_{\text{Na,300K}}$. This technique has hitherto remained unexplored in the realm of energy materials design and presents a timely and pragmatic opportunity to identify NASICON-type solid electrolytes demonstrating excellent $\sigma_{\text{Na,300K}}$ performance, as finding easily-accessible but effective descriptors. Significantly, our model facilitates a comprehensive understanding within the expansive landscape of NASICON-type compositions, providing valuable insights into the specific conditions under which the desired $\sigma_{\text{Na,300K}}$ would be achieved.

**Sampling protocol.** In this study, we constructed the sampling protocol to build an explicit regression model of $\sigma_{\text{Na,300K}}$ for NASICON-type solid electrolytes, as represented in Fig. 1. We initially collected 19 samples as diverse as possible that represent various types of $\text{Na}_n M_m M'_{m'} \text{Si}_{3-p-a} \text{P}_p \text{As}_a \text{O}_{12}$ for the training dataset of the regression modelling: quaternary, quinary, and senary, and $v(M') = 2$, 3, and 4; $v(M')$ is the valence of metal ions $M'$. The protocol comprises four parts: structure search, data training, model test, and model revision for experimental values.

**Traning dataset**

Type I: quaternary
Type II: quinary and $M$ = Zr
Type III: senary, $M$ = Zr, and $v(M')$ = 2
Type IV: quinary, $M$ = Zr, and $v(M')$ = 3
Type V: senary, $M$ = Zr, and $v(M')$ = 3
Type VI: senary, $M$ = Zr, and $v(M')$ = 4

**Structure search**

Ewald summation sampling
↓
Geometry-optimization with DFT

**Data training**

Single-$T$ "long-time" diagnoses
  $\Delta\tau$ = 1 fs, $\tau$ = 1 ns, and $T$ = 300 K
Multivariate beta regression modelling for $D_{Na,300K}$
  Found descriptors: $d_1$ and $<d_{Na-Na}>$
Demensionality reduction for $\sigma_{Na,300K}$

$\sigma_{Na,300K}$ — $\rho_{Na}$, $D_{Na,300K}$, $d_1$, $<d_{Na-Na}>$, $n$, $<r_M>$, $V$

Identify competitive NASICON with $n$ and $<r_M>$
Na$_{2.75}$Zr$_{1.75}$Nb$_{0.25}$Si$_2$PO$_{12}$: $\sigma_{Na,300K,sim}$ > 10$^{-3}$ S·cm$^{-1}$

**Model test**

Multi-$T$ DFT-MD calculations
  $\Delta\tau$ = 1 fs, $\tau$ = 600 ps,
  and $T$ = 300, 500, 700, and 900 K
Na$_{2.75}$Zr$_{1.75}$Nb$_{0.25}$Si$_2$PO$_{12}$: $\sigma_{Na,300K}$ > 10$^{-3}$ S·cm$^{-1}$

**Model revision for experimental values**

Introduce probit function
$R^2$ = 0.718 for $n_{data}$ = 140

**Fig. 1** Sampling protocol to build an explicit regression model of $\sigma_{Na,300K}$ for NASICON-type solid electrolytes. Given the representative samples for various types of NASCION compositions Na$_n$M$_m$M$'_{m'}$Si$_{3-p-a}$P$_p$As$_a$O$_{12}$, the protocol comprises four parts: structure search, data training, model test, and model revision for experimental values. For the denotations, the main text is referred to.

For the structure search, we utilized Ewald summation sampling[54] and density functional theory (DFT) to optimize the crystal structures of the 19 representative in-silico samples for $Na_nM_mM'_{m'}Si_{3-p-a}P_pAs_aO_{12}$, starting from the monoclinic structure $Na_3Zr_2Si_2PO_{12}$ with $C2/c$ symmetry that was determined experimentally.[11] Subsequently, we performed DFT-MD calculations for these 19 training samples at room temperature ($T = 300$ K), which we call single-$T$ "long-time" diagnoses (with a time step of $\Delta\tau = 1$ fs over a simulation time of $\tau = 1$ ns).[10] Based on the obtained data, we exhaustively explored multivariate beta regression models,[55-57] incorporating 17 features encompassing electrostatic, diffusion-pathway, and geometrical characteristics. Here, the adoption of beta regression modelling was employed to effectively distinguish between low- and high-performance samples in a manner similar to binary classification. The selection of the optimal model was based on the highest value of the pseudo-goodness-of-fit metric $R^2_{pseudo}$.[56] Herein, we found two primary descriptors; the bottleneck width along the diffusion paths for Na-ions $d_1$ and the average Na-Na distance $\langle d_{Na-N} \rangle$ play pivotal roles in modulating the room-temperature Na-ion self-diffusion coefficients $D_{Na,300K}$. Then, we introduced two secondary descriptors easily accessible from the chemical formula only; the Na-ion content $n$ as of $Na_nM_mM'_{m'}Si_{3-p-a}P_pAs_aO_{12}$ affect $d_1$, $\langle d_{Na-Na}\rangle$, and Na-ion density $\rho_{Na}$, and the average $\langle r_M \rangle$ of ionic radii $r_M$ for metal ions $M$ and $M'$ (excluding Na-ions)[58] affect $d_1$ and $\langle d_{Na-Na}\rangle$. Based on the beta regression model above, the easily-accessible descriptors $n$, $n^2$, and $\langle r_M \rangle$ allow for the final explicit form of regression model.

To elucidate the viability of this regression model, we selected the composition $Na_{2.75}Zr_{1.75}Nb_{0.25}Si_2PO_{12}$, hitherto unexplored, which was identified as a potential candidate for $\sigma_{Na,300K,sim} > 10^{-3}$ S·cm$^{-1}$ ($\sigma_{Na,300K,sim}$ is the simulated value of $\sigma_{Na,300K}$ by using the regression model). It exhibited thermodynamic phase stability, that is, low decomposition energy above the convex hull ($E_{\text{hull}} = 6.49$ meV·atom$^{-1}$). To estimate not only $\sigma_{\text{Na,300K}}$ but also the Na-ion bulk activation energy $E_a$, we performed additional DFT-MD calculations for the test sample at multiple temperatures ($T = 300, 500, 700,$ and 900 K), which we call multi-$T$ DFT-MD calculations (with $\Delta \tau = 1$ fs and $\tau = 600$ ps).[10] The results confirmed the validity of the model: $\sigma_{\text{Na,300K}} = 1.45 \times 10^{-3}$ S·cm$^{-1}$ and $E_a = 232$ meV. Then, we extended the regression modelling for the experimental values of $\sigma_{\text{Na,300K}}$. The adjusted model, wherein the probit function of the aforementioned model was introduced, demonstrated the capability to fit 140 experimental values with the goodness-of-fit metric $R^2 = 0.718$.

**Building a regression model.** Table 1 presents the values of $E_{hull}$, $\sigma_{Na,300K}$, the room-temperature Na-ion self-diffusion coefficients $D_{Na,300K}$, and the $R$-squared values $R^2_{MSD}$ for the mean squared displacement (MSD) curves regressed against sampled time intervals $\Delta\tau_{MSD}$ through the single-$T$ "long-time" diagnoses, given the fixed unit cells optimized by using DFT. $E_{hull}$ across the 19 samples showed the structural (meta)stability ($E_{hull} < 40$ meV·atom$^{-1}$). Among the total dataset of $n_{data} = 19$, five samples were classified as low-performance, characterized by $\sigma_{Na,300K} < 10^{-4}$ S·cm$^{-1}$ and $D_{Na,300K} < 10^{-9}$ cm$^2$·s$^{-1}$. These low-performance samples were predominantly observed in the quaternary types. The remaining eight samples were categorized as high-performance, exhibiting $\sigma_{Na,300K} > 10^{-4}$ S·cm$^{-1}$ and $D_{Na,300K} > 10^{-9}$ cm$^2$·s$^{-1}$. In Supplementary Fig. 1, we represent the MSD curves for the 19 samples. Meanwhile, the pristine composition $Na_3Zr_2Si_2PO_{12}$ exhibited $\sigma_{Na,300K} = 1.09 \times 10^{-3}$ S·cm$^{-1}$, which aligns with the experimental measurement of the bulk ion conductivity $\sigma_{Na,300K} \cong 10^{-3}$ S·cm$^{-1}$ as justifying the quality of the data training method.[59]

**Table 1** Values for energies $E_{hull}$ above the convex hull, room-temperature Na-ion conductivities $\sigma_{Na,300K}$, room-temperature Na-ion self-diffusion coefficients $D_{Na,300K}$, and $R$-squared values $R^2_{MSD}$ for the MSD curves regressed against sampled time intervals $\Delta\tau_{MSD}$, which were estimated by the single-$T$ "long-time" diagnoses (with $\Delta\tau = 1$ fs and $\tau = 1$ ns at $T = 300$ K) for the 19 samples.

| Compounds | $E_{hull}$ (meV·atom$^{-1}$) | $\sigma_{Na,300K}$ (S·cm$^{-1}$) | $D_{Na,300K}$ (cm$^2$·s$^{-1}$) | $R^2_{MSD}$ |
|---|---|---|---|---|
| $NaTi_2P_3O_{12}$ | 20.4 | $5.47 \times 10^{-7}$ | $2.10 \times 10^{-11}$ | 0.0500 |
| $NaGe_2P_3O_{12}$ | 16.5 | $5.55 \times 10^{-7}$ | $1.94 \times 10^{-11}$ | 0.0464 |
| $NaZr_2P_3O_{12}$ | 0 | $3.18 \times 10^{-6}$ | $1.36 \times 10^{-10}$ | 0.308 |
| $NaZr_2As_3O_{12}$ | 0 | $1.65 \times 10^{-6}$ | $7.85 \times 10^{-11}$ | 0.219 |
| $Na_3In_2P_3O_{12}$ | 37.0 | $5.88 \times 10^{-3}$ | $8.48 \times 10^{-8}$ | 0.859 |
| $Na_3Zr_2Si_2PO_{12}$ | 0 | $1.09 \times 10^{-3}$ | $1.62 \times 10^{-8}$ | 0.920 |
| $Na_{3.25}Zr_{1.875}Mg_{0.125}Si_2PO_{12}$ | 0.578 | $5.99 \times 10^{-4}$ | $8.17 \times 10^{-9}$ | 0.881 |
| $Na_{3.75}Zr_{1.625}Mg_{0.375}Si_2PO_{12}$ | 5.42 | $6.15 \times 10^{-5}$ | $7.26 \times 10^{-10}$ | 0.347 |
| $Na_{3.25}Zr_{1.875}Ca_{0.125}Si_2PO_{12}$ | 1.76 | $1.10 \times 10^{-4}$ | $1.51 \times 10^{-9}$ | 0.143 |
| $Na_{3.5}Zr_{1.75}Ca_{0.25}Si_2PO_{12}$ | 4.14 | $1.07 \times 10^{-4}$ | $1.37 \times 10^{-9}$ | 0.0549 |
| $Na_{3.75}Zr_{1.625}Zn_{0.375}Si_2PO_{12}$ | 17.1 | $1.23 \times 10^{-3}$ | $1.46 \times 10^{-8}$ | 0.767 |
| $Na_{1.5}Zr_{1.5}Al_{0.5}P_3O_{12}$ | 9.32 | $1.48 \times 10^{-3}$ | $4.06 \times 10^{-8}$ | 0.942 |
| $Na_{1.5}Zr_{1.5}Sc_{0.5}P_3O_{12}$ | 6.46 | $1.74 \times 10^{-3}$ | $4.97 \times 10^{-8}$ | 0.977 |
| $Na_{1.5}Zr_{1.5}Ga_{0.5}P_3O_{12}$ | 16.1 | $9.03 \times 10^{-4}$ | $2.52 \times 10^{-8}$ | 0.912 |
| $Na_{2.875}Zr_{0.125}In_{1.875}P_3O_{12}$ | 36.1 | $1.71 \times 10^{-3}$ | $2.57 \times 10^{-8}$ | 0.943 |
| $Na_{3.375}Zr_{1.625}Sc_{0.375}Si_2PO_{12}$ | 33.6 | $1.53 \times 10^{-3}$ | $2.01 \times 10^{-8}$ | 0.661 |
| $Na_3Zr_{1.25}Ti_{0.75}Si_2PO_{12}$ | 17.6 | $2.08 \times 10^{-3}$ | $2.98 \times 10^{-8}$ | 0.642 |
| $Na_3Zr_{1.75}Ti_{0.25}Si_2PO_{12}$ | 7.13 | $1.51 \times 10^{-3}$ | $2.21 \times 10^{-8}$ | 0.928 |
| $Na_3Zr_{1.75}Sn_{0.25}Si_2PO_{12}$ | 8.13 | $3.95 \times 10^{-3}$ | $5.81 \times 10^{-8}$ | 0.989 |

While some of the high-performance samples displayed reliable MSD curves with high $R^2_{\text{MSD}}$ values, the others yielded low $R^2_{\text{MSD}}$ values due to the insufficient occurrence of site-to-site jump events given the low temperature $T = 300$ K. In Supplementary Fig. 1, we additionally illustrate the finite quantities of trajectory samples exhibiting squared displacements surpassing the squared average Na-Na distances $\langle d_{\text{Na-Na}} \rangle^2$, which are related to the sizes of $D_{\text{Na,300K}}$ [see Supplementary Fig. 2]. Consequently, the MSD curves suffered from considerable noise, leading to lower reliability of the $D_{\text{Na,300K}}$ and $\sigma_{\text{Na,300K}}$ values. To address the uncertainties associated with $D_{\text{Na,300K}}$, multivariate beta regression modelling was employed to clearly differentiate between low- and high-performance samples as well as to elucidate the descriptors with significant contribution towards $D_{\text{Na,300K}}$ prediction. The beta regression response function based on Eq. (12) (described in the Method section) was given by

$$\eta = 0.261 \log_{10} D_{\text{Na,300K}} + 2.82. \quad (1)$$

In the present regression modelling, a total of 17 features (descriptor candidates) were included, comprising two electrostatic, eight diffusion-pathway, and seven geometrical features (see Supplementary Note 1 for details and Supplementary Table 2 for the feature dataset). Among the various fitted models, the one whose sigmoid function response $\bar{\eta}$ fits $\eta$ [see Eqs. (10)—(12) in the Method section] with the highest $R^2_{\text{pseudo}}$ (see Supplementary Note 2) is composed of two $z$-scored primary descriptors as follows:

$$h = 1.28 \, z[d_1] - 1.26 \, z[\langle d_{\text{Na-Na}} \rangle] + 0.719, \quad (2)$$

with the common precision $\varphi = 7.49$ (see Supplementary Note 2): $R^2_{\text{pseudo}} = 0.797$ (see Fig. 2). The Pearson correlation coefficient between $d_1$ and $\langle d_{\text{Na-N}} \rangle$ was given as $r(d_1, \langle d_{\text{Na-N}} \rangle) =$

.204 (see Supplementary Note 2), indicating the absence of the significant multicollinearity issue within Eq. (2). To further evaluate whether these two descriptors are indeed predictive, we repeatedly compute $R^2_{\text{pseudo}}$ but with the omission of 4 random samples ($n_{\text{data}} = 15$; $n_{\text{data}}$ is the number of data) for 100 times. The mean and the standard deviation for the $R^2_{\text{pseudo}}$ values were given as 0.803 and 0.0368, respectively, indicating its robustness. Then, from Eqs. (2), (10), and (11), we defined $\sigma_{\text{Na,300K,sim}}$ as a function of $d_1$ and $\langle d_{\text{Na-Na}} \rangle$ by taking $z[d_1] = (d_1 - 1.66\ \text{Å})/0.0882\ \text{Å}$ and $z[\langle d_{\text{Na-Na}} \rangle] = (\langle d_{\text{Na-Na}} \rangle - 3.64)/0.164$ and assuming $\bar{\eta} \cong \eta$ and $\sigma_{\text{Na,300K,sim}} \cong \sigma_{\text{Na,300K}} = \frac{(z_{\text{Na}} F)^2 \rho_{\text{Na}}}{RT} D_{\text{Na,300K}}$, where $z_{\text{Na}}$ ($= +1$) is the valence for a Na-ion, and $F$ and $R$ denote the Faraday constant and the gas constant, respectively:

$$\sigma_{\text{Na,300K,sim}}(d_1, \langle d_{\text{Na-Na}} \rangle) \equiv$$

$$(6.20 \times 10^6\ \text{S} \cdot \text{cm}^{-1})\left(\rho_{\text{Na}}\ \text{Å}^3\right) 10^{\left(\frac{1}{1+e^{(-14.5\ \text{Å}^{-1} d_1 + 7.70 \langle d_{\text{Na-Na}} \rangle - 4.527)}} - 2.82\right)/0.261}. \quad (3)$$

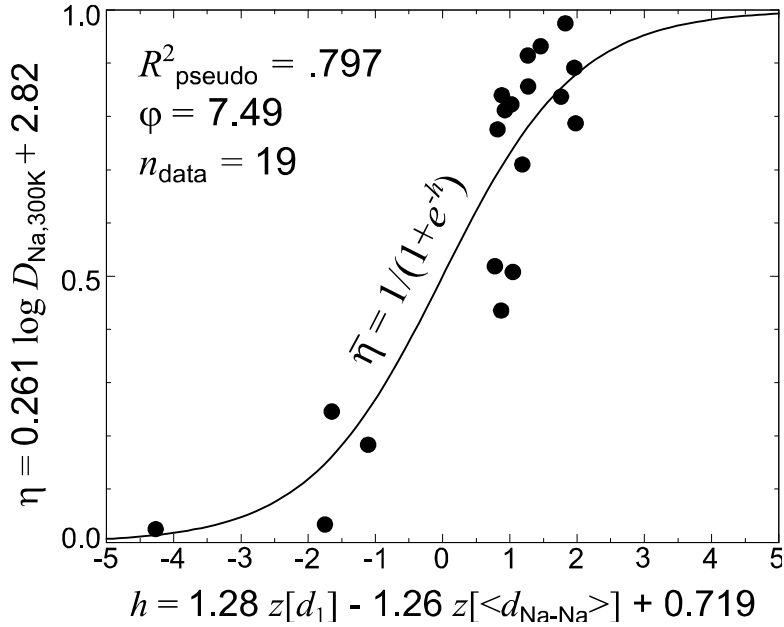

**Fig. 2** Multivariate beta regression models with the highest pseudo-goodness-of-fit $R^2_{\text{pseudo}}$ value given two descriptors allowed, wherein low- and high-performance samples are easily distinguishable in a binary-like manner. For the notations, Eqs. (2), (10)—(12) are referred to.

In previous studies, diffusion-pathway descriptors, including $d_1$, have been identified as crucial indicators that contribute to decreased steric hindrance during ion migrations.[60, 61] Meanwhile, the decrease in $\langle d_{\text{Na-}}\rangle$ may imply the importance of Coulombic repulsion among Na-ions. Consequently, it can be inferred that a diffusion pathway characterized by decreased steric hindrance and shorter interstitial distances among Na-ions would result in the increase of $D_{\text{Na,300K}}$. In Supplementary Note 3, we compare the descriptors for $D_{\text{Na,300K}}$ in NASICONs and Na-ion sulfides.[10]

To construct a regression model that is readily interpretable, we further found secondary easily-accessible descriptors for $\sigma_{\text{Na,300K}}$ ($\propto \rho_{\text{Na}} D_{\text{Na,300K}}$) which involved transitioning from the primary descriptors $\rho_{\text{Na}}$, $d_1$, and $\langle d_{\text{Na-N}} \rangle$. (Herein, the term "secondary" descriptors signify information given independently of structure, whereas "primary" descriptors pertain to the Na-ion diffusion pathways inherent in the NASICON structure.) For the purpose, we developed linear regression models as described in Supplementary Note 4, wherein $\rho_{\text{Na}}$, $d_1$, and $\langle d_{\text{Na-Na}} \rangle$ were regressed with $n$ and $\langle r_{\text{M}} \rangle$. In broad terms, it can be stated that $n$ plays a pivotal role in the modulation of both $\rho_{\text{Na}}$ and $\langle d_{\text{Na-Na}} \rangle$, whereas $\langle r_{\text{M}} \rangle$ exerts significant influence over $d_1$. With Supplementary Eqs. (S6)—(S9), $\sigma_{\text{Na,300K,sim}}$ was rewritten as a function of $n$ and $\langle r_{\text{M}} \rangle$ from Eq. (3):

$$\sigma_{\text{Na,300K,sim}}(n, \langle r_{\text{M}} \rangle) \equiv$$

$$(6.20 \times 10^6 \text{ S} \cdot \text{cm}^{-1}) (0.00350n + 0.000527) \, 10^{\left(\frac{1}{1+\exp(8.02-5.86n+1.13n^2-5.22 \text{ Å}^{-1} \langle r_{\text{M}} \rangle)} - 2.82\right)/0.261},$$

(4)

wherein the exponential part ($\propto D_{\text{Na,300K}}$) would reach its zenith approximately $n = 2.59$. We plot $\sigma_{\text{Na,300K,sim}}$ against $n$ for different values of $\langle r_{\text{M}} \rangle$ in the range of $[0.3, \ 0.5]$ Å in Fig. 3a. We estimated the overall goodnesses-of-fit for $\log_{10} \sigma_{\text{Na,300K,sim}}$ against the trained dataset of $\log_{10} \sigma_{\text{Na,300K}}$ (given in the single-$T$ "long-time" diagnoses; listed in Table 1). The model exhibits a reasonable $R^2$ value of 0.694. It should be briefly noted that the beta regression model involving $n$ and $\langle r_{\text{M}} \rangle$ only yielded the small size of $R^2_{\text{pseudo}} = 0.243$, which can be attributed to the omission of nonlinear terms such as $n^2$.

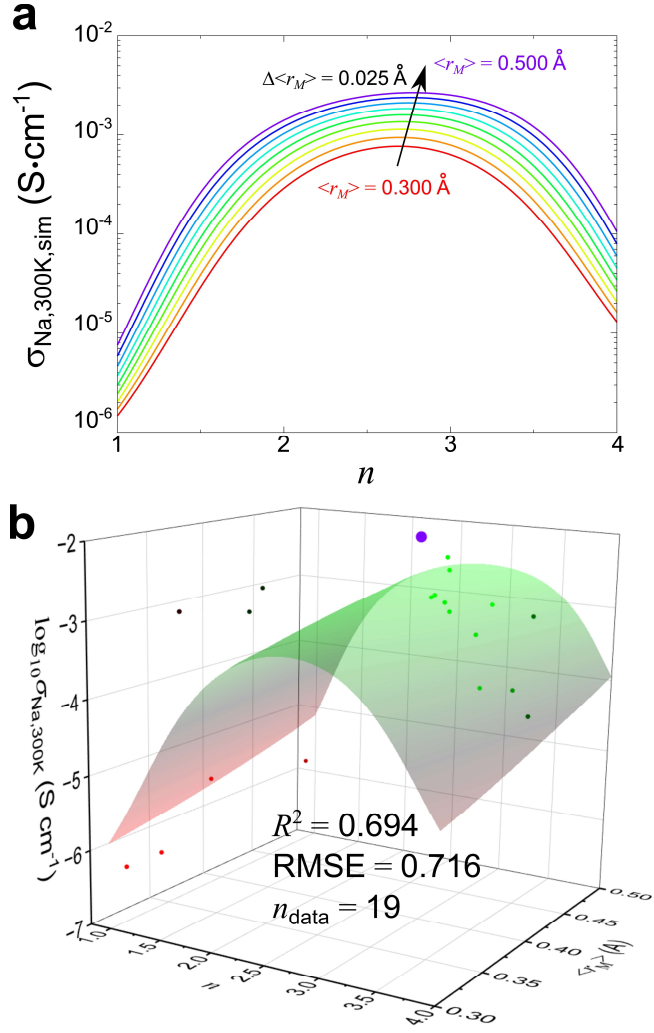

**Fig. 3 a** Simulated room-temperature Na-ion conductivity $\sigma_{\text{Na,300K,sim}}$ [see Eq. (4)] with varying Na-content $n$ for given $\langle r_M \rangle$. Hue-color-scaled $\langle r_M \rangle$ are given in the range of [0.3, 0.5] Å with increment of 0.025 Å. **b** $\sigma_{\text{Na,300K,sim}}$ (represented by curved surface) compared with theoretical values $\sigma_{\text{Na,300K}}$ listed in Table 1 (small dots) trained through the single-$T$ "long-time" diagnoses (with $\Delta\tau = 1$ fs and $\tau = 1$ ns at $T = 300$ K). The values for $R^2$ and root mean square errors (RMSE) are also represented for $\log_{10} \sigma_{\text{Na,300K}} = \log_{10} \sigma_{\text{Na,300K,sim}}$. The large purple dot represents the test case of $\text{Na}_{2.75}\text{Zr}_{1.75}\text{Nb}_{0.25}\text{Si}_2\text{PO}_{12}$ through the single-$T$ "long-time" diagnosis: $\sigma_{\text{Na,300K}} = 1.00 \times 10^{-2}$ S·cm$^{-1}$.

**Promising composition.** We take a quinary chemical formula $Na_nZr_{2-m'}M'_{m'}Si_{3-p}P_pO_{12}$ as our target, where the charge neutrality $[m' = (4 - p - n)/\{v(M') - 4\}]$ imposes a constraint for $\langle r_M \rangle$:

$$\langle r_M \rangle = \frac{1}{5}\left\{\left(2 - \frac{4-p-n}{v(M')-4}\right)r_{Zr} + \frac{4-p-n}{v(M')-4}r_{M'} + (3-p)r_{Si} + pr_P\right\}; \quad (5)$$

it is given that $r_{Zr} = 0.72$ Å, $r_{Si} = 0.26$ Å, and $r_P = 0.17$ Å.[58] Thus, when a specific value of $p$ is chosen alongside a particular selection of $M'$ that yield $v(M')$ and $r_{M'}$, $\sigma_{Na,300K,sim}$ can be simplified into a function of $n$. Under $p = 1$ as an example, we illustrate the variation of $\sigma_{Na,300K,sim}$ concerning different values of $n$ for various $M'$ (24 in total) through Figs. 4a, 4b, 4c, and 4d. For $p = 1$, $v(M') = 4$, and $n = 3$ [as presented in Fig. 3c], it is given that

$$\sigma_{Na,300K,sim}(m', r_{M'}) \equiv (1.06 \times 10^{-6} \text{ S} \cdot \text{cm}^{-1}) \exp\left[\frac{1}{0.113 + 0.0224\exp(0.751m' - 1.04 \text{ Å}^{-1} r_{M'}m')}\right].$$

(6)

While Deng et al. previously proposed that the maximal value of $\sigma_{Na,300K}$ is anticipated around at $n = 2.4$,[62] we further notice that the optimization of $\sigma_{Na,300K,sim}$ does not exclusively hinge on $n$ considering the diverse array of $M'$ selections.

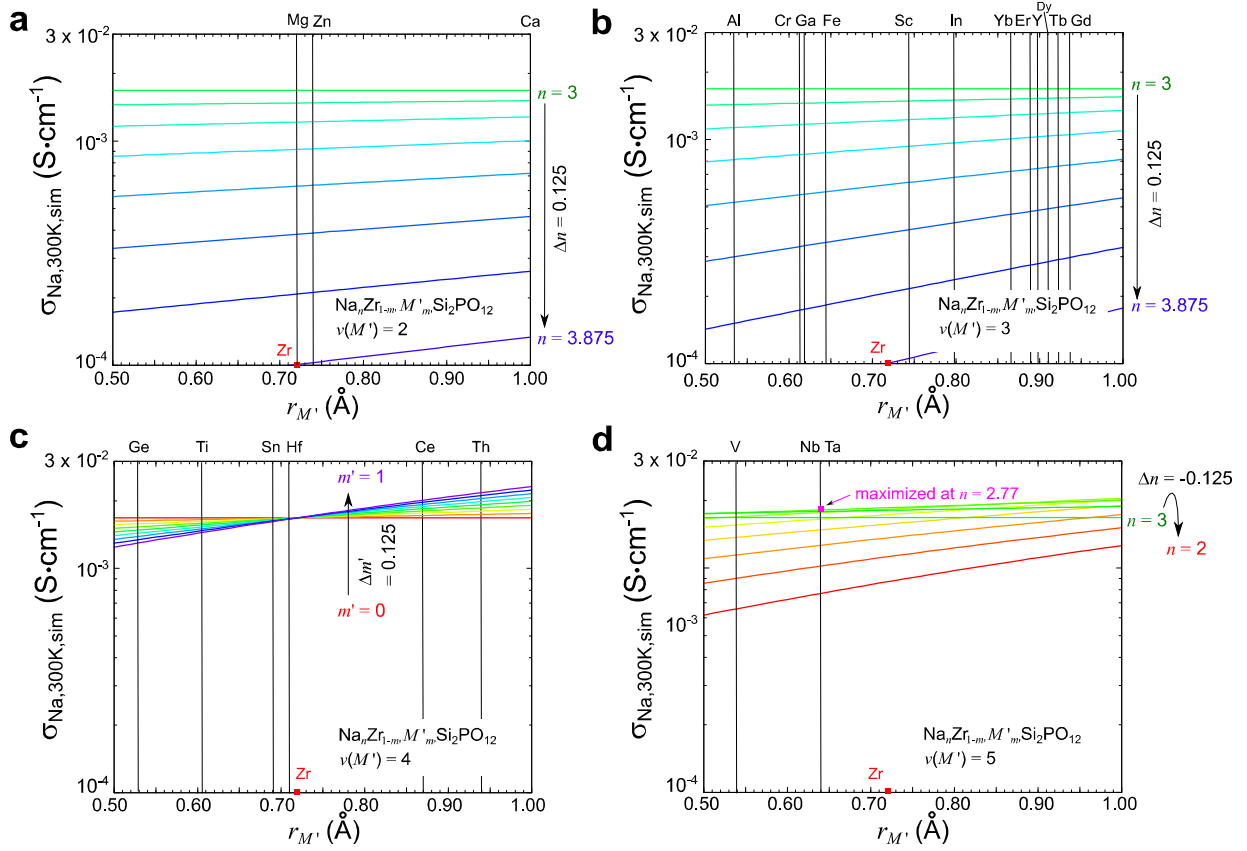

**Fig. 4** $\sigma_{Na,300K,sim}$ varying with the Na-ion content $n$ (or the dopant content $m'$: hue-color-scaled) and the ionic radius $r_{M'}$ of a dopant ion $M'$ for $Na_nZr_{2-m'}M'_{m'}Si_2PO_{12}$ given **a** divalent [$v(M') = 2$], **b** trivalent [$v(M') = 3$], **c** tetravalent [$v(M') = 4$], and **d** pentavalent [$v(M') = 5$] $M'$: 24 types of $M'$ in total. $r_{M'}$ for various types of $M'$ are represented by the vertical lines, and $r_{Zr}$ is also denoted by red dots.[72] The pink dot in **d** represents the case of the maximized $\sigma_{Na,300K,sim}$ across all $M'$ except for Ce and Th: $M' = $ Nb and Ta and $n = 2.77$.

Herein, we focused on a selection of potential $M'$ candidates while addressing two key considerations. First, among the 24 choices, $M' = $ Nb, Ce, Ta, and Th were found competitive as depicted in Figs. 3c and 3d. Second, building upon the findings by Ouyang et al.,[63] $r_{M'} \cong r_{Zr} = $ 0.72 Å would offer notable advantages in terms of structural stability and synthesis accessibility. Considering $r_{\text{Nb}} = 0.64$ Å, $r_{\text{Ce}} = 0.87$ Å, $r_{\text{Ta}} = 0.64$ Å, and $r_{\text{Th}} = 0.94$ Å, $M' = $ Nb and Ta emerge as particularly intriguing, with which $\sigma_{\text{Na,300K,sim}}$ would be maximized around at $n = 2.77$: $\sigma_{\text{Na,300K,sim}} = 1.82 \times 10^{-3}$ S·cm$^{-1}$. Subsequently, we formulated a test composition, namely $\text{Na}_{2.75}\text{Zr}_{1.75}\text{Nb}_{0.25}\text{Si}_2\text{PO}_{12}$, with the anticipation that the scenario would follow a similar trend in the case of Ta. For the structure searches, we carried out Ewald summation sampling and geometry optimizations using DFT for $\text{Na}_{2.75}\text{Zr}_{1.75}\text{Nb}_{0.25}\text{Si}_2\text{PO}_{12}$, considering not only the monoclinic structure but also the rhombohedral structure.[11] Starting from the monoclinic configuration, $\text{Na}_{2.75}\text{Zr}_{1.75}\text{Nb}_{0.25}\text{Si}_2\text{PO}_{12}$ exhibited $E_{\text{hull}} = 6.49$ meV·atom$^{-1}$ indicating its high structural stability. Moreover, the free energy calculated through DFT for the monoclinic structures was found more stable by 3.52 meV·atom$^{-1}$ compared to the rhombohedral counterpart. Fig. 5a illustrate the crystal structure of $\text{Na}_{2.75}\text{Zr}_{1.75}\text{Nb}_{0.25}\text{Si}_2\text{PO}_{12}$ based on the initial monoclinic structure.

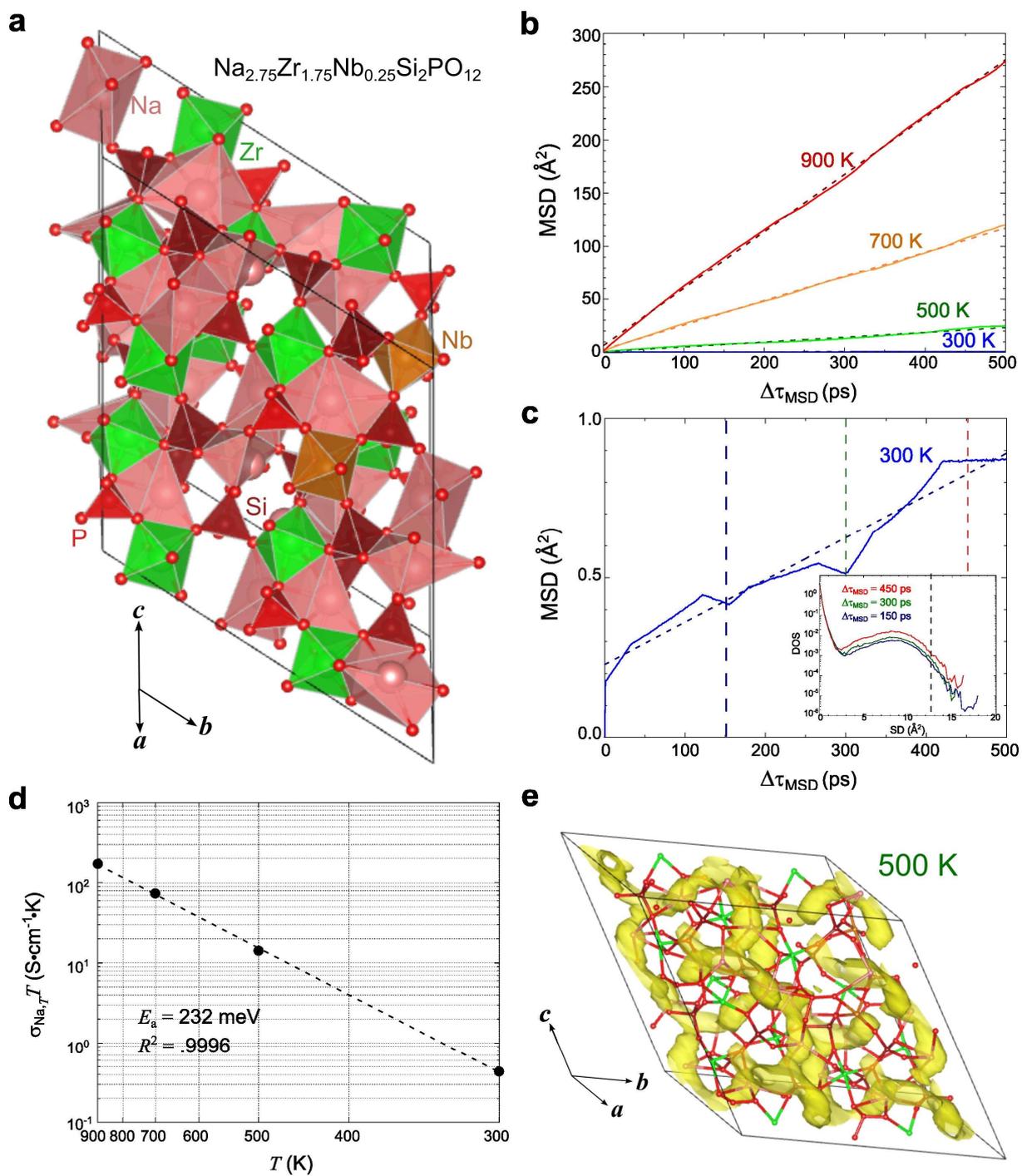

**Fig. 5 a** Crystal structure for $Na_{2.75}Zr_{1.75}Nb_{0.25}Si_2PO_{12}$. Pink, green, gold, dark red, and red polyhedra denote $NaO_x$, $ZrO_6$, $NbO_6$, $SiO_4$, and $PO_4$, respectively. The black arrows represent the lattice vectors. **b** The mean squared displacement (MSD) curves against sampled time intervals $\Delta\tau_{MSD}$ given by the multi-$T$ DFT-MD calculations (with $\Delta\tau = 1$ fs and $\tau = 600$ ps at $T = 300$, 500, 700, and 900 K). **c** The magnified view of the MSD curve for $T = 300$ K. In the inset, the density of states (DOS) for the trajectory samples, that is, squared displacement (SQ) data points across ensembles and times, where the area is normalized to 1. The blue, green, and red vertical dashed lines denote $\Delta\tau_{MSD} = 150, 300$, and $450$ ps, respectively, which correspond to the DOS curves of the same colors in the inset. The black vertical dashed line in the inset denotes SD = $\langle d_{Na-Na}\rangle^2$, the squared Na-Na distance. In **b** and **c**, the dashed lines with slopes represent regressions against sampled time intervals $\Delta\tau_{MSD}$. **d** The Arrhenius plot in the $\sigma_{Na,T} T$ -$T$ domain with the calculated value of the Na-ion bulk activation energy $E_a$ and the $R^2$ values. **e** The trajectory density plot at $T = 500$ K represented by yellow isosurfaces.

We proceeded with multi-$T$ DFT-MD calculations to investigate $\sigma_{Na,T}$ and $E_a$ given the thermally-equilibrated unit cell volumes during the $NpT$ pre-treatment stage. We represent the MSD curves in Figs. 5b and 5c. As indicated by Fig. 5c, there was the finite quantity of trajectory samples exhibiting squared displacements surpassing $\langle d_{Na-N}\rangle^2$ at $T = 300$ K. In Fig. 5d, the corresponding Arrhenius plot in the $\sigma_{Na,T} T$ -$T$ domain is represented, wherein $\sigma_{Na,300K}$ and $E_a$ were estimated as $1.45 \times 10^{-3}$ S·cm$^{-1}$ and 232 meV, respectively. It is noteworthy that $\sigma_{Na,T}$ with $T = 500, 700$, and 900 K exhibit a nearly identical extrapolated value to that of $\sigma_{Na,300K}$ given $R^2 = 0.9996$. This observation serves to partially validate the utilization of room-temperature DFT-MD calculations, particularly for high-performance samples. In Fig. 5e, we present the trajectory density plot given in $T = 500$ K, wherein the migration paths for Na-ions are intricately interconnected within the bulk. The estimated values of $\sigma_{\text{Na},T}$, $D_{\text{Na},T}$, and $R^2_{\text{MSD}}$ are also shown in Table 2.

**Table 2** Values of Na-ion conductivities $\sigma_{\text{Na},T}$, self-diffusion coefficients $D_{\text{Na},T}$, and $R$-squared values $R^2_{\text{MSD}}$ for the MSD curves regressed against sampled time intervals $\Delta\tau_{\text{MSD}}$, which were estimated by the multi-$T$ DFT-MD calculations (with $\Delta\tau$ = 1 fs and $\tau$ = 600 ps at $T$ = 300, 500, 700, and 900 K) for $Na_{2.75}Zr_{1.75}Nb_{0.25}Si_2PO_{12}$.

| $T$ (K) | $\sigma_{\text{Na},T}$ (S·cm$^{-1}$) | $D_{\text{Na},T}$ (cm$^2$·s$^{-1}$) | $R^2_{\text{MSD}}$ |
|---|---|---|---|
| 300 | $1.45 \times 10^{-3}$ | $2.21 \times 10^{-8}$ | .950 |
| 500 | $2.96 \times 10^{-2}$ | $7.55 \times 10^{-7}$ | .978 |
| 700 | $1.05 \times 10^{-1}$ | $3.83 \times 10^{-6}$ | .999 |
| 900 | $1.92 \times 10^{-1}$ | $8.96 \times 10^{-6}$ | .999 |

Furthermore, to ensure a fair and consistent comparison with the values of $\sigma_{\text{Na},300\text{K}}$ and $D_{\text{Na},300\text{K}}$ presented in Table 1, which were determined through the single-$T$ "long-time" diagnoses based on fixed unit cells optimized by using DFT without undergoing thermal equilibration, we carried out the same procedure for $Na_{2.75}Zr_{1.75}Nb_{0.25}Si_2PO_{12}$. Unexpectedly, the results yielded $\sigma_{\text{Na},300\text{K}} = 1.00 \times 10^{-2}$ S·cm$^{-1}$, visually depicted by the large purple dot in Fig. 3b. This outcome needs careful consideration, and its comprehensive discussion is given in detail in Supplementary Note 5.

**Model revision for experimental cases.** We carried out investigations to ascertain whether the easily-accessible descriptors $n$, $n^2$, and $\langle r_M \rangle$ are adequate for accurately fitting the experimental values of $\sigma_{Na,300K}$ as well. We meticulously collected all available experimental data points to our best knowledge, amounting to $n_{data} = 182$, which are exhaustively listed in Supplementary Table 3.[15, 19, 20, 22, 23, 28-31, 34, 37, 41, 43-45, 47, 48, 53] We then refined the dataset by excluding compositions with $Cr^{3+}$, $Fe^{3+}$, $Ce^{3+}$, $Gd^{3+}$, and $Yb^{3+}$, where $d$ or $f$ electrons are partially filled. This exclusion narrowed down the dataset to $n_{data} = 140$. In our pursuit of capturing not only the high but also the low conductivity regime $\sigma_{Na,300K} < 10^{-6}$ S·cm$^{-1}$, which remains challenging to reconcile with Eq. (4), we introduced the probit function $p\sigma$ into our analytical framework:

$$\log_{10} p\sigma(n, \langle r_M \rangle) \equiv a_1 \log_{10}\left[\frac{1}{2}\left\{1 + \text{erf}\left(\frac{a_2 + \sum_{j=1}^{df(j)} b_j [a_3 \log_{10} \sigma_{Na,300K,sim}(n, \langle r_M \rangle) + a_4]^{c_j}}{\sqrt{2}}\right)\right\}\right] + a_5, \quad (7)$$

where $df(j)$ represents the count of power functions of $\sigma_{Na,300K,sim}(n, \langle r_M \rangle)$, namely Eq. (4), incorporated into the model, and $a_1$, $a_2$, $a_3$, $a_4$, and $a_5$ serve as normalization factors for the error function; $b_j$ and $c_j$ are the only free parameters to the model, given $df(j)$. As $df(j)$ increases [$df(j) = 1$, $2$, and $3$], the overall goodnesses-of-fit against the experimental values of $\log_{10} \sigma_{Na,300K}$ were found as $R^2 = 0.687, 0.718$, and $0.733$, respectively. At $df(j) = 2$, where $R^2$ approaches saturation, the coefficients in Eq. (7) were specified as follows: $a_1 = 12.6$, $a_2 = -3.01$, $a_3 = 0.308$, $a_4 = 1.79$, $a_5 = -14.8$, $(b_1, c_1) = (3.00, 0.0235)$, and $(b_2, c_2) = (1.30, 1.40)$. We plot $p\sigma(n, \langle r_M \rangle)$ with $df(j) = 2$ against $n$ for different values of $\langle r_M \rangle$ in the range of [0.3, 0.5] Å in Fig. 6a. The revised model exhibits a reasonable $R^2$ value of 0.718, even when confronted with diverse synthesis techniques, distinct space groups, grain boundary effects, ion-ion correlated effects, and other pertinent factors, as illustrated in Fig. 6b. This observation leads us to the conclusion that the development of a model for $\sigma_{\text{Na,300K}}$ incorporating $n$, $n^2$, and $\langle r_{\text{M}} \rangle$ is indeed feasible.

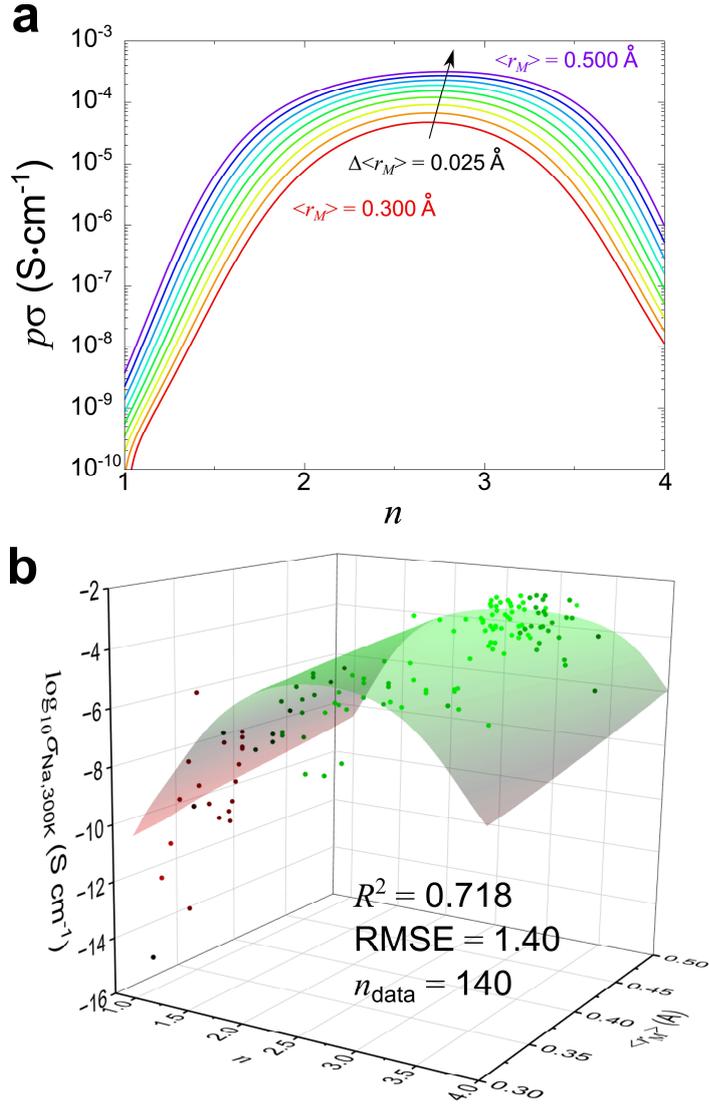

**Fig. 6 a** Simulated room-temperature Na-ion conductivity $p\sigma(n, \langle r_M \rangle)$ [see Eq. (7)] with varying Na-content $n$ for given $\langle r_M \rangle$. Hue-color-scaled $\langle r_M \rangle$ are given in the range of [0.3, 0.5] Å with increment of 0.025 Å. **b** $p\sigma(n, \langle r_M \rangle)$ (represented by curved surface) compared with experimental values $\sigma_{Na,300K}$ listed in Supplementary Table 3 (dots), illustrated with easily-accessible descriptors $n$ and $\langle r_M \rangle$. The values for $R^2$ and root mean square errors (RMSE) are also represented for $\log_{10} \sigma_{Na,300K} = \log_{10} p\sigma_{Na,300K,sim}$. Herein, compositions with $Cr^{3+}$, $Fe^{3+}$, $Ce^{3+}$, $Gd^{3+}$, and $Yb^{3+}$ are excluded, where $d$ or $f$ electrons are partially filled: $n_{data} = 140$.

**Summary and Outlook**

This study employed DFT-MD to establish an explicit regression model for predicting $\sigma_{\text{Na,300K}}$ within NASICON-type solid electrolytes for potential use in solid-state batteries. By using regression techniques including multivariate beta regression modelling, we successfully built a model with two easily-accessible descriptors only: $n$ (and $n^2$ as well) and $\langle r_\text{M} \rangle$. This simplicity in the model's features suggests an efficient and resource-effective approach to predicting ion conductivity, making it potentially applicable to a wide range of materials. We also note that a closely related study by Wang et al., briefly discussed that the finding optimal value of $\sigma_{\text{Na,300K}}$ may be possible with descriptors similar to ours, albeit rather in broad strokes.[64]

This model led to the exploration of $\text{Na}_{2.75}\text{Zr}_{1.75}\text{Nb}_{0.25}\text{Si}_2\text{PO}_{12}$, as well. The material was found thermodynamically stable with $E_{\text{hull}} = 6.49$ meV·atom$^{-1}$. Subsequent multi-$T$ DFT-MD calculations confirmed that this promising yet unexplored stable composition $\text{Na}_{2.75}\text{Zr}_{1.75}\text{Nb}_{0.25}\text{Si}_2\text{PO}_{12}$ may likely achieve $\sigma_{\text{Na,300K}} > 10^{-3}$ S·cm$^{-1}$. The result demonstrates the potential of this material as solid electrolytes in Na-ion-based solid-state batteries. Also, given the same ionic radii for Nb$^{5+}$ and Ta$^{5+}$ (0.64 Å),[58] the case of $M' = $ Ta with the same Na-ion content is also worth consideration for the future study. Meanwhile, its revision successfully predicted the 140 experimental values of $\sigma_{\text{Na,300K}}$ as well, demonstrating the robustness and applicability of the model to real-world scenarios.

**Associated content**

Supporting Information

The details of multivariate beta regression models and considered features (descriptor candidates), the mean squared displacement plots given by the single-temperature "long-time" diagnoses, the feature (descriptor candidates) dataset used for the multivariate beta regression modelling, the comparison of descriptors for $D_{\text{Na,300K}}$ in NASICONs and Na-ion sulfides, the linear regressed models for the primary descriptors against the secondary ones, the experimental data for $\sigma_{\text{Na,300K}}$ in the literature, and the results given in the single-temperature "long-time" diagnosis for $\text{Na}_{2.75}\text{Zr}_{1.75}\text{Nb}_{0.25}\text{Si}_2\text{PO}_{12}$.

**Author information**

**Acknowledgments**

This research was supported in part by MEXT as "Program for Promoting Research on the Supercomputer Fugaku" grant number JPMXP1020200301, Data Creation and Utilization Type Material Research and Development Project grant number JPMXP1121467561 and Materials Processing Science project ("Materealize") grant number JPMXP0219207397, and by JSPS KAKENHI grant numbers JP21K14729, as well as JST through ALCA-SPRING grant number JPMJAL1301, GteX grant numbers JPMJGX23S2, and COI-NEXT grant number JPMJPF2016. The calculations were performed on the supercomputers at NIMS (Numerical Materials Simulator). Visualization for crystal structures was made with the VESTA software.[65] Plots were generated using gnuplot 5.4.[66]

**Methods**

**Site arrangement sampling.** Using EwaldSolidSolution,[54] we performed Ewald summation sampling starting from the monoclinic structure $Na_3Zr_2Si_2PO_{12}$ with $C2/c$ symmetry that was determined experimentally.[11] By expanding the primitive cell, we created a $1 \times 2 \times 2$ supercell $Na_{32}^+Zr_{16}^{4+}Si_{16}^{4+}P_8^{5+}O_{96}^{2-}$, considering the presence of 8 excess Na-ion sites identified by a large Debye-Waller factor. Here, the lattice constants were given as $a = 9.0062$ Å, $b = 18.012$ Å, $c = 18.410$ Å, $\alpha = \beta = 61.308°$, and $\gamma = 60.168°$. While maintaining the number of O-ion sites (96), we randomly generated 957,600 to 3,783,780 site arrangements for each case of $Na_n M_m M'_{m'} Si_{3-p-a} P_p As_a O_{12}$ by modifying $Na_{32}^+Zr_{16}^{4+}Si_{16}^{4+}P_8^{5+}O_{96}^{2-}$ to $Na_{8n}^+ M_{8m} M'_{8m'} Si_{24-8p-8a}^{4+} P_{8p}^{5+} As_{8a}^{5+} O_{96}^{2-}$, wherein the total number of ion sites ranged from 144 to 166.

**Geometry optimization.** Following the site arrangement sampling, the five most stable site arrangement samples for each case were subjected to DFT geometry optimizations using the Vienna Ab Initio Simulation Package (VASP). We employed the generalized gradient approximation (GGA) and the projector augmented wave (PAW) method basis set.[67-71] The geometry optimizations included both site positions and lattice constants. Monkhorst-Pack $\boldsymbol{k}$-grids were set at $2 \times 2 \times 2$,[72] and the kinetic energy cutoff of 520 eV was used. Convergence criteria of $< 0.01$ eV·Å$^{-1}$ for forces and $< 10^{-5}$ eV·atom$^{-1}$ for energy were applied. Some pseudopotentials included semicore electrons as valence states for specific elements: Ca, Sc, and Zr (semicore $s$ electrons); Na, Mg, Ti, Nb, and Ta ($p$); and Ga, Ge, In, and Sn ($d$). For the other elements, standard pseudopotential forms were employed. Then, the lowest-energy structure sample for each investigated composition was selected for subsequent DFT-MD calculations, wherein we calculated $E_{\text{hull}}$ for all the samples by using the Computational Phase Diagram App provided by MaterialsProject.org[73, 74] to verify their thermodynamic (meta)stability.

**DFT-MD for data training.** The single-$T$ "long-time" diagnoses were carried out at $T = 300$ K given the geometry-optimized cell structures described above. First, a total of 10,000 DFT-MD steps (10 ps) were performed to ensure thermal equilibrations by using the Nosé-Hoover thermostat ($NVT$ ensemble) implemented in VASP.[75, 76] Subsequently, DFT-MD production runs were executed for trajectory sampling over $\tau = 1$ ns ($NVT$). Throughout the DFT-MD calculations, $\Delta \tau = 1$ fs, a $1 \times 1 \times 1$ $\boldsymbol{k}$-grid (that is, Γ only), and a kinetic energy cutoff of 400 eV were employed. The pseudopotentials were used in their standard forms except for Ca and Zr (with semicore $s$ electrons) and Nb ($p$), and the calculations were performed using the GGA and the PAW method basis set.[67-71]

From the sampled trajectories, the Na-ion self-diffusion coefficients $D_{\text{Na},T} = M_s/(2d)$ at $T$ were estimated by carrying out regression analyses on the diffusive (linear) regime of the mean squared displacement (MSD) curves against sampled time intervals $\Delta \tau_{\text{MSD}}$, up to $\Delta \tau_{\text{MSD}} = 800$ ps; $D_{\text{Na},T}$ was obtained as the slope $M_s$ of the MSD-$\Delta \tau_{\text{MSD}}$ regression line at $T$, considering the three-dimensional nature of Na-ion diffusion ($d = 3$). We also estimated $R$-squared values $R^2_{\text{MSD}}$ for the MSD curves regressed against $\Delta \tau_{\text{MSD}}$. Then, the Na-ion ionic conductivity $\sigma_{\text{Na},T}$ at $T$ is estimated by using the Nernst-Einstein equation

$$\sigma_{\text{Na},T} = \frac{(z_{\text{Na}}F)^2 \rho_{\text{Na}}}{RT} D_{\text{Na},T}, \quad (9)$$

where $z_{\text{Na}}$ ($= +1$) is the valence for a Na-ion, $\rho_{\text{Na}}$ is the Na-ion density, and $F$ and $R$ denote the Faraday constant and the gas constant, respectively.

**Multivariate beta regression modelling.** The goal is to find a beta regression model $h$ whose sigmoid function response $\bar{\eta}$ well-fits the performance scores $\eta$ for $\log_{10} D_{\text{Na,300K}}$:

$$h = c + \sum_{i=1}^{df} c(x_i) z[x_i], \tag{10}$$

$$\bar{\eta} = 1/(1 + e^{-h}), \tag{11}$$

and

$$\eta = \frac{1}{n_{\text{data}}} \left\{ \frac{\log_{10} D_{\text{Na,300K}} - \text{mi}\,[\log_{10} D_{\text{Na,300K}}]}{\max[\log_{10} D_{\text{Na,300K}}] - \min[\log_{10} D_{\text{Na,300K}}]} (n_{\text{data}} - 1) + \frac{1}{2} \right\}, \tag{12}$$

where $x_i$, $z[x_i]$, and $c(x_i)$ ($i$ = integer), $c$, $df$, and $n_{\text{data}}$ denote the independent variables (that is, features considered as descriptor candidates), their $z$-scored values, and their coefficients, the constant term, the maximum number of taken $x_i$, i.e., the degree of freedom, and the number of data for $\log_{10} D_{\text{Na,300K}}$, respectively. Eq. (4) was taken in that the target dependent variable in beta regression analysis is scaled in the range of (0,1), as introduced first by Smithson and Verkuilen.[77]

While we leave the detailed descriptions of the maximum likelihood estimation and the pseudo-goodnees-of-fit $R^2_{\text{pseudo}}$ for the exhaustive search of the possible multivariate beta regression models in Supplementary Note 2, we briefly note that the R-package betareg was used for the analyses,[57, 78, 79] and that the Zeo++ package was used for the three descriptors $d_0$, $d_1$, and $V_p$ by referring to the Shannon ionic radii for ion sites.[58, 80, 81]

**DFT-MD for test dataset.** The multi-$T$ DFT-MD calculations were carried out at $T = $ 300, 500, 700, and 900 K. First, a total of 40,000 DFT-MD steps (40 ps) were performed to achieve thermal and volume equilibrations by using the Langevin thermostat with the Parinello-Rahman algorithm (*NpT* ensemble) implemented in VASP.[82, 83] During this process, the averaged lattice constants were calculated over the last 10,000 DFT-MD steps (30 - 40 ps) to account for the thermally induced cell volume expansion. Subsequently, with the averaged lattice constants, thermal equilibration runs were repeated for 10,000 DFT-MD steps (10 ps) under the Nosé-Hoover thermostat (*NVT*). Finally, product runs were carried out afterwards for trajectory sampling over $\tau = 600$ ps (*NVT*). Meanwhile, the choices of $\Delta\tau$, the $\boldsymbol{k}$-grid, the kinetic energy cutoff, and the pseudopotentials and the post-process for $D_{\text{Na},T}$ and $\sigma_{\text{Na},T}$ were common to those of the DFT-MD for data training. $D_{\text{Na},T}$ were estimated by carrying out regression analyses on MSD curves against sampled time intervals $\Delta\tau_{\text{MSD}}$, up to $\Delta\tau_{\text{MSD}} = 500$ ps.

Supplementary Information

# Predicting room-temperature conductivity of Na-ion super ionic conductors with the minimal number of easily-accessible descriptors

*Seong-Hoon Jang,*[1,2*] *Randy Jalem*[2]*, and Yoshitaka Tateyama*[2,3]

[1] Institute for Materials Research, Tohoku University, 2-1-1 Katahira, Aoba-ku, Sendai, 980-8577, Japan

[2] Research Center for Energy and Environmental Materials (GREEN), National Institute for Materials Science (NIMS), 1-1 Namiki, Tsukuba, Ibaraki 305-0044, Japan

[3] Laboratory for Chemistry and Life Science, Tokyo Institute of Technology, 4259 Nagatsuta, Midori-ku, Yokohama, 226-8501, Japan

*Corresponding author: jang.seonghoon.b4@tohoku.ac.jp

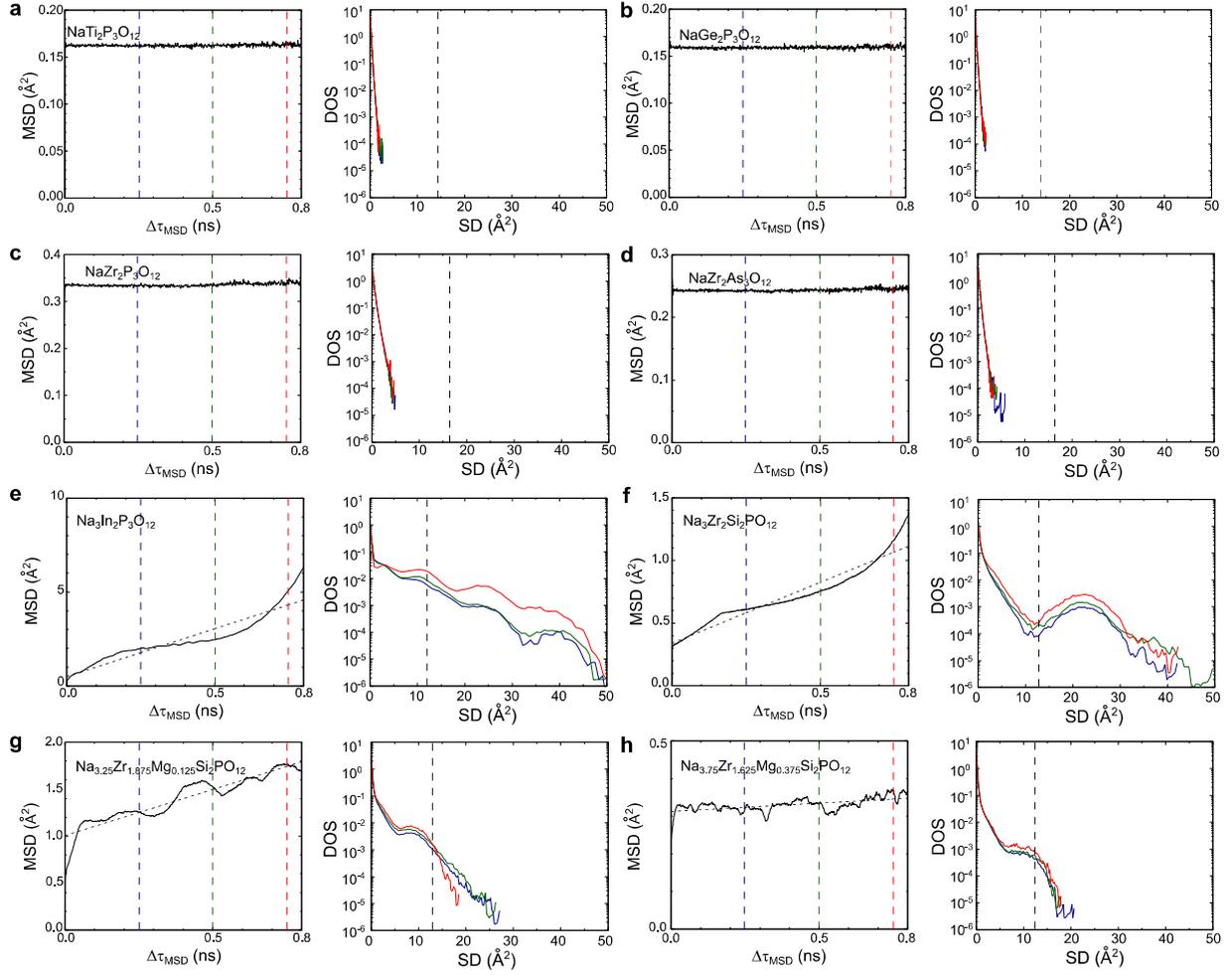

**Supplementary Fig. 1** Mean squared displacement (MSD) curves (left) and the density of states (DOS) for the trajectory samples, that is, squared displacement (SQ) data points across ensembles and times (right). They were given by single-$T$ "long-time" diagnoses (with $\Delta\tau = 1$ fs and $\tau = 1$ ns at $T = 300$ K). In the left plot for each case, the black dashed line represents regressions against sampled time intervals $\Delta\tau_{MSD}$, and the blue, green, and red vertical dashed lines denote $\Delta\tau_{MSD} = 0.25, 0.50,$ and $0.75$ ns, respectively, which correspond to the DOS curves of the same colors in the right plot for each case. The black vertical dashed line in the right plot for each case denotes $SD = \langle d_{Na-Na} \rangle^2$, the squared Na-Na bond distance. The area for each DOS plot is normalized to 1. For $\langle d_{Na-Na} \rangle$, refer to Supplementary Table 2.

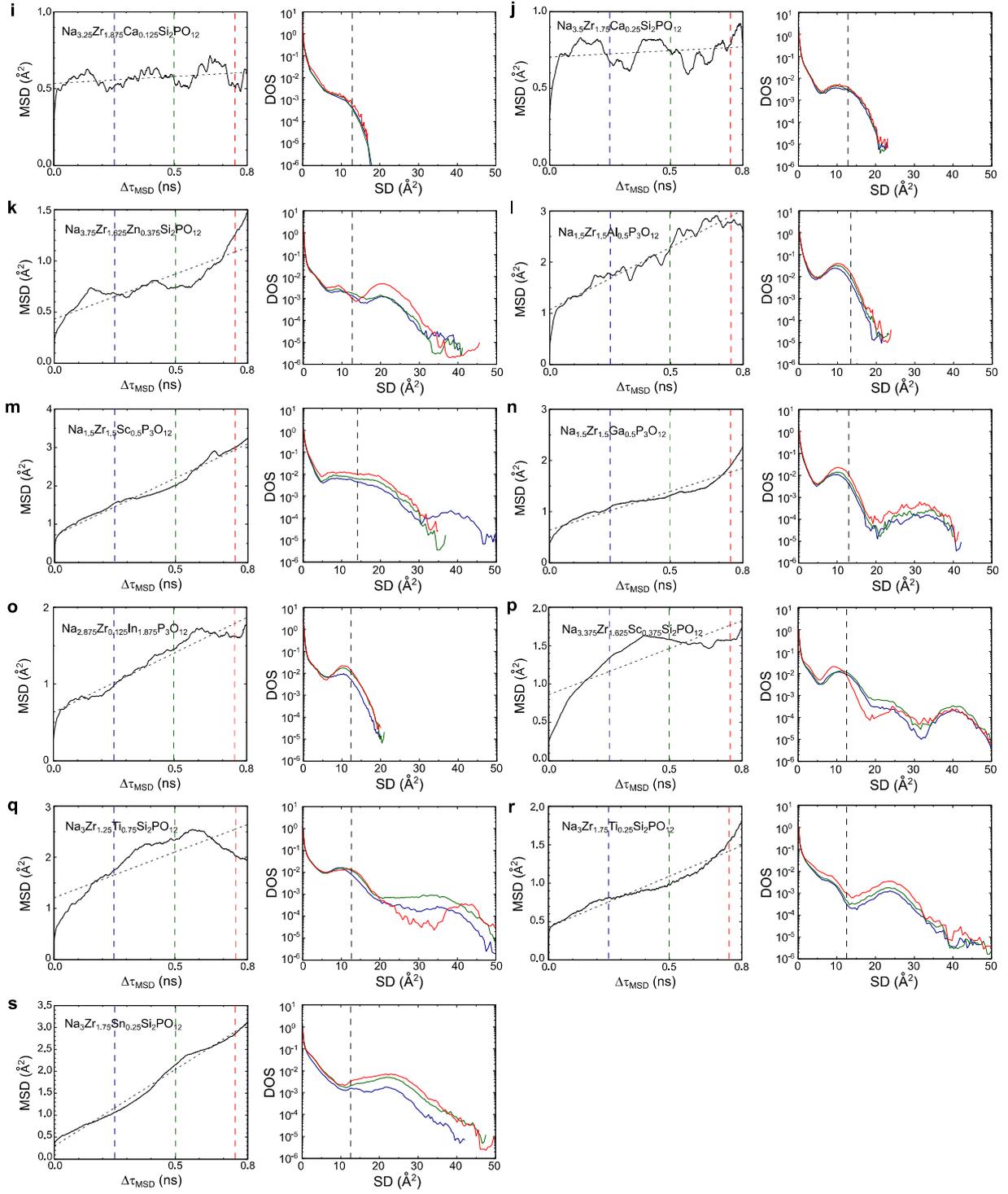

(to be continued)

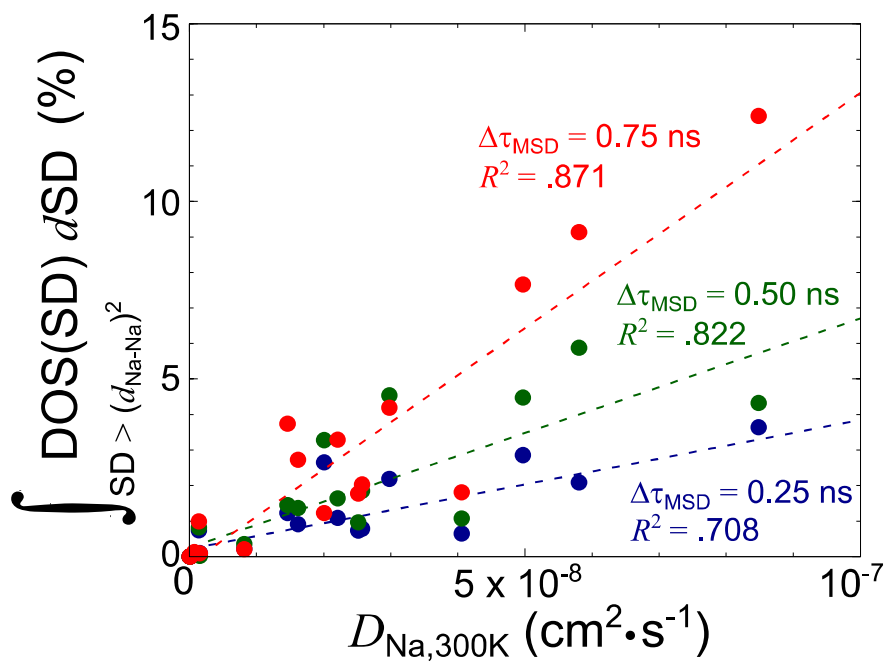

**Supplementary Fig. 2** DOS areas with SD > $\langle d_{Na-Na}\rangle^2$ are plot against the room-temperature Na-ion self-diffusion coefficients $D_{Na,300K}$ calculated from the MSD curves at $\Delta\tau_{MSD}$ = 0.25, 0.50, and 0.75 ns. For $\langle d_{Na-Na}\rangle$, refer to Supplementary Table 2.

**Supplementary Note 1** Considered features (descriptor candidates) for the room-temperature Na-ion self-diffusion coefficients $D_{\mathrm{Na,300K}}$

**Supplementary Table 1** Detailed descriptions of considered features (descriptor candidates) for the room-temperature Na-ion self-diffusion coefficient $D_{\mathrm{Na,300K}}$.

| | |
|---|---|
| Electrostatic feature | |
| $\langle C_\mathrm{P} \rangle$ | the average charge for polyhedra $MO_y$ ($M$ = Ti, Ge, Zr, Mg, Zn, Al, Sc, Ga, In, Si, P, and As); the valence (-2) of O were divided by the number of surrounding metal ions since neighboring $MO_y$ are not completely separated. |
| $\langle \chi_M \rangle$ | average of electronegativity $\chi_M$ for metal ions (excluding Na-ions) |
| Diffusion-pathway features | |
| $d_0$ | the broadest width along the diffusion paths for Na-ions |
| $d_1$ | the bottleneck width along the diffusion paths for Na-ions |
| $V_\mathrm{p}$ | the porous volume given to a Na-ion |
| $\rho_\mathrm{Na}$ | Na-ion density |
| $n$ | Na-ion content as of $\mathrm{Na}_n M_m M'_{m'} \mathrm{Si}_{3-p-a} \mathrm{P}_p \mathrm{As}_a \mathrm{O}_{12}$ |
| $\langle d_\mathrm{Na-Na} \rangle$ | the average Na-Na bond length |
| $\mathrm{med}(d_\mathrm{Na-Na})$ | the median Na-Na bond length |
| $\langle r_\mathrm{M} \rangle$ | the average of ionic radii $r_M$ for metal ions (excluding Na-ions) |
| Geometrical features | |
| $\langle d_\mathrm{Na-O} \rangle$ | the average Na-O bond length |
| $\langle n_\mathrm{Na-O} \rangle$ | the average coordination number for a Na-ion to O-ions |
| Site-specific local geometrical features (especially for polyhedra $\mathrm{NaO}_x$) | |
| $\langle V_{\mathrm{NaO}_x} \rangle$ | the average volume of polyhedra $\mathrm{NaS}_x$ |
| $\langle \min(\Omega_{\mathrm{NaO}_3}) \rangle$ | the average narrowest Na-3O solid angle for polyhedra $\mathrm{NaO}_x$ |
| $\langle \Omega_{\mathrm{NaO}_3} \rangle$ | the average Na-3O solid angle across for polyhedra $\mathrm{NaO}_x$ |
| $\langle \max(\Omega_{\mathrm{NaO}_3}) \rangle$ | the average widest Na-3O solid angle for polyhedra $\mathrm{NaO}_x$ |
| $\mathrm{stdev}(\Omega_{\mathrm{NaO}_3})$ | the standard deviation of the Na-3O solid angles for polyhedra $\mathrm{NaO}_x$ |

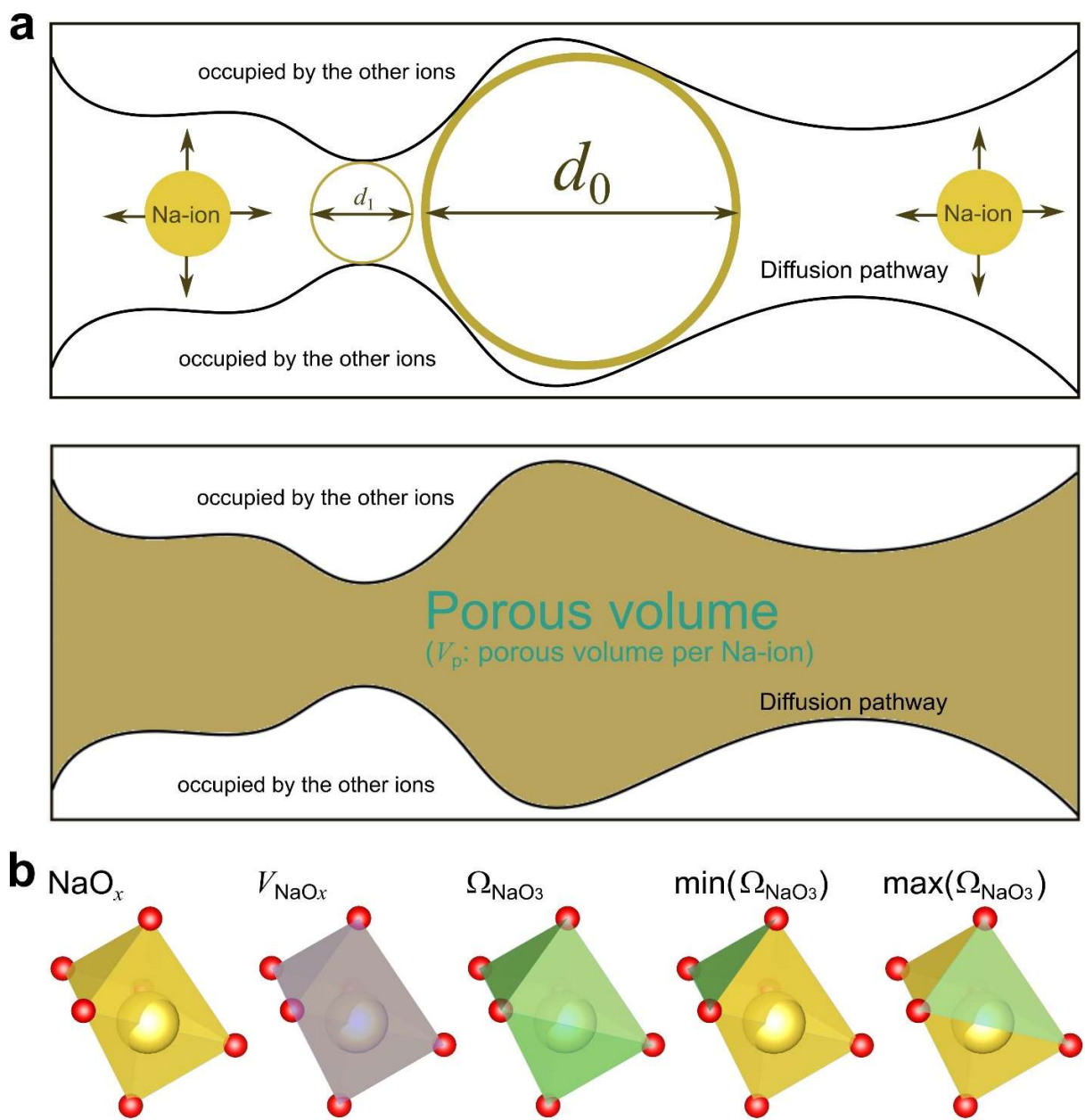

**Supplementary Fig. 3** Schematic illustrations of **a** diffusion-pathway features $d_0$, $d_1$, and $V_p$ and **b** site-specific local geometrical features $V_{NaO_x}$, $\Omega_{NaO_3}$, $\min(\Omega_{NaO_3})$, and $\max(\Omega_{NaO_3})$. For the definitions, Supplementary Table 1 is referred to.

The descriptions of the 17 features $x_i$ are provided in Supplementary Table 1. We also illustrate the diffusion-pathway features $d_0$, $d_1$, and $V_p$ and site-specific local geometrical features $V_{NaO_x}$, $\Omega_{NaO_3}$, $\min(\Omega_{NaO_3})$, and $\max(\Omega_{NaO_3})$ in Supplementary Fig. 3. To obtain the estimates for $x_i = d_0, d_1$, and $V_p$, a Voronoi tessellation technique was employed on the given sample cell space after manually excluding Na-ions. By using the Voronoi nodes obtained from the tessellation, $d_0$ represents the diameter of the largest sphere centered on a node that touches the surface of the remaining ions. It corresponds to the widest section along the diffusion path. Meanwhile, $d_1$ corresponds to the diameter of the largest sphere that can move along the nodes, representing the narrowest section along the diffusion path. Additionally, $V_p$ refers to the probe-occupiable volume per Na-ion. Since Na-ions were excluded from the sample cell, a probe with the size of a Na-ion ionic radius (1.0 Å)[3] can move freely within the void. In the analysis, the Shannon ionic radii for the considered ion sites were referred to,[1] and the Zeo++ package was utilized for the calculations.[2,3]

**Supplementary Table 2** Feature dataset for the 19 samples given in Table 1. The definitions for the descriptors are given in Supplementary Table 1. The average $\langle r_M \rangle$ of ionic radii $r_M$ for metal ions and the volume $V$ for a unit cell of each sample are also presented. In the last column, we added the case of $Na_{2.75}Zr_{1.75}Nb_{0.25}Si_2PO_{12}$.

| Descriptor | $NaTi_2P_3O_{12}$ | $NaGe_2P_3O_{12}$ | $NaZr_2P_3O_{12}$ | $NaZr_2As_3O_{12}$ |
|---|---|---|---|---|
| $\langle C_P \rangle$ | -0.184 | -0.184 | -0.184 | -0.184 |
| $\langle \chi_M \rangle$ | 1.93 | 2.12 | 1.85 | 1.84 |
| $d_0$ (Å) | 2.38 | 2.33 | 2.60 | 2.51 |
| $d_1$ (Å) | 1.57 | 1.37 | 1.73 | 1.75 |
| $V_p$ (Å$^3$) | 3.00 | 2.46 | 20.6 | 40.0 |
| $\rho_{Na}$ (Å$^{-3}$) | 0.00421 | 0.00462 | 0.00378 | 0.00340 |
| $n$ | 1 | 1 | 1 | 1 |
| $\langle d_{Na-Na} \rangle$ (Å) | 3.78 | 3.73 | 4.06 | 4.03 |
| $med(d_{Na-Na})$ (Å) | 3.78 | 3.73 | 4.06 | 4.03 |
| $\langle d_{Na-O} \rangle$ (Å) | 2.49 | 2.47 | 2.57 | 2.55 |
| $\langle n_{Na-O} \rangle$ | 5.07 | 5.17 | 4.93 | 4.74 |
| $\langle V_{NaO_x} \rangle$ (Å$^3$) | 16.7 | 16.5 | 17.7 | 17.2 |
| $\langle min(\Omega_{NaO_3}) \rangle$ | 0.631 | 0.569 | 0.656 | 0.655 |
| $\langle \Omega_{NaO_3} \rangle$ | 1.48 | 1.40 | 1.57 | 1.57 |
| $\langle max(\Omega_{NaO_3}) \rangle$ | 2.15 | 2.41 | 2.09 | 2.08 |
| $stdev(\Omega_{NaO_3})$ | 0.592 | 0.691 | 0.568 | 0.573 |
| $\langle r_M \rangle$ (Å) | 0.344 | 0.314 | 0.390 | 0.489 |
| $V$ (Å$^3$) | 1900 | 1730 | 2120 | 2350 |

| Descriptor | $Na_3In_2P_3O_{12}$ | $Na_3Zr_2Si_2PO_{12}$ | $Na_{3.25}Zr_{1.875}Mg_{0.125}Si_2PO_{12}$ | $Na_{3.75}Zr_{1.625}Mg_{0.375}Si_2PO_{12}$ |
|---|---|---|---|---|
| $\langle C_P \rangle$ | -0.579 | -0.622 | -0.583 | -0.694 |
| $\langle \chi_M \rangle$ | 2.03 | 1.73 | 1.73 | 1.73 |
| $d_0$ (Å) | 2.47 | 2.54 | 2.53 | 2.54 |
| $d_1$ (Å) | 1.65 | 1.72 | 1.67 | 1.63 |
| $V_p$ (Å$^3$) | 9.02 | 10.1 | 9.28 | 7.40 |
| $\rho_{Na}$ (Å$^{-3}$) | 0.0112 | 0.0109 | 0.0118 | 0.0137 |
| $n$ | 3 | 3 | 3.25 | 3.75 |
| $\langle d_{Na-Na} \rangle$ (Å) | 3.47 | 3.57 | 3.59 | 3.55 |
| med($d_{Na-Na}$) (Å) | 3.47 | 3.51 | 3.57 | 3.51 |
| $\langle d_{Na-O} \rangle$ (Å) | 2.56 | 2.58 | 2.57 | 2.57 |
| $\langle n_{Na-O} \rangle$ | 5.64 | 5.58 | 5.34 | 5.10 |
| $\langle V_{NaO_x} \rangle$ (Å$^3$) | 20.9 | 18.8 | 17.1 | 16.3 |
| $\langle \min(\Omega_{NaO_3}) \rangle$ | 0.714 | 0.604 | 0.605 | 0.551 |
| $\langle \Omega_{NaO_3} \rangle$ | 1.35 | 1.40 | 1.50 | 1.53 |
| $\langle \max(\Omega_{NaO_3}) \rangle$ | 2.72 | 2.71 | 3.25 | 3.43 |
| stdev($\Omega_{NaO_3}$) | 0.647 | 0.806 | 0.926 | 0.972 |
| $\langle r_M \rangle$ (Å) | 0.422 | 0.426 | 0.426 | 0.426 |
| $V$ (Å$^3$) | 2150 | 2200 | 2200 | 2200 |

| Descriptor | $Na_{3.25}Zr_{1.875}Ca_{0.125}Si_2PO_{12}$ | $Na_{3.5}Zr_{1.75}Ca_{0.25}Si_2PO_{12}$ | $Na_{3.75}Zr_{1.625}Zn_{0.375}Si_2PO_{12}$ | $Na_{1.5}Zr_{1.5}Al_{0.5}P_3O_{12}$ |
|---|---|---|---|---|
| $\langle C_P \rangle$ | -0.583 | -0.639 | -0.722 | -0.216 |
| $\langle \chi_M \rangle$ | 1.72 | 1.71 | 1.75 | 1.71 |
| $d_0$ (Å) | 2.58 | 2.57 | 2.56 | 2.59 |
| $d_1$ (Å) | 1.64 | 1.65 | 1.63 | 1.76 |
| $V_p$ (Å$^3$) | 9.53 | 8.67 | 7.51 | 9.99 |
| $\rho_{Na}$ (Å$^{-3}$) | 0.0118 | 0.0126 | 0.0136 | 0.00587 |
| $n$ | 3.25 | 3.5 | 3.75 | 1.5 |
| $\langle d_{Na-Na} \rangle$ (Å) | 3.59 | 3.57 | 3.55 | 3.66 |
| med($d_{Na-Na}$) (Å) | 3.57 | 3.59 | 3.49 | 3.61 |
| $\langle d_{Na-O} \rangle$ (Å) | 2.58 | 2.60 | 2.57 | 2.54 |
| $\langle n_{Na-O} \rangle$ | 5.48 | 5.45 | 5.25 | 5.29 |
| $\langle V_{NaO_x} \rangle$ (Å$^3$) | 18.1 | 18.3 | 16.5 | 17.3 |
| $\langle \min(\Omega_{NaO_3}) \rangle$ | 0.638 | 0.562 | 0.582 | 0.626 |
| $\langle \Omega_{NaO_3} \rangle$ | 1.47 | 1.47 | 1.53 | 1.45 |
| $\langle \max(\Omega_{NaO_3}) \rangle$ | 3.07 | 3.42 | 3.37 | 2.50 |
| stdev($\Omega_{NaO_3}$) | 0.884 | 0.978 | 0.949 | 0.712 |
| $\langle r_M \rangle$ (Å) | 0.433 | 0.440 | 0.428 | 0.318 |
| $V$ (Å$^3$) | 2210 | 2230 | 2200 | 2050 |

| Descriptor | $Na_{1.5}Zr_{1.5}Sc_{0.5}P_3O_{12}$ | $Na_{1.5}Zr_{1.5}Ga_{0.5}P_3O_{12}$ | $Na_{2.875}Zr_{0.125}In_{1.875}P_3O_{12}$ | $Na_{3.375}Zr_{1.625}Sc_{0.375}Si_2PO_{12}$ |
|---|---|---|---|---|
| $\langle C_P \rangle$ | -0.216 | -0.216 | -0.514 | -0.639 |
| $\langle \chi_M \rangle$ | 1.85 | 1.89 | 2.01 | 1.73 |
| $d_0$ (Å) | 2.69 | 2.59 | 2.58 | 2.55 |
| $d_1$ (Å) | 1.77 | 1.73 | 1.62 | 1.62 |
| $V_p$ (Å$^3$) | 15.5 | 10.1 | 9.31 | 9.28 |
| $\rho_{Na}$ (Å$^{-3}$) | 0.00564 | 0.00579 | 0.0107 | 0.0123 |
| $n$ | 1.5 | 1.5 | 2.875 | 3.375 |
| $\langle d_{Na-Na} \rangle$ (Å) | 3.77 | 3.62 | 3.52 | 3.54 |
| $med(d_{Na-Na})$ (Å) | 3.81 | 3.62 | 3.50 | 3.53 |
| $\langle d_{Na-O} \rangle$ (Å) | 2.55 | 2.55 | 2.54 | 2.57 |
| $\langle n_{Na-O} \rangle$ | 4.95 | 5.40 | 5.39 | 5.21 |
| $\langle V_{NaO_x} \rangle$ (Å$^3$) | 16.4 | 17.7 | 18.4 | 17.8 |
| $\langle \min(\Omega_{NaO_3}) \rangle$ | 0.568 | 0.605 | 0.787 | 0.667 |
| $\langle \Omega_{NaO_3} \rangle$ | 1.54 | 1.44 | 1.45 | 1.49 |
| $\langle \max(\Omega_{NaO_3}) \rangle$ | 2.74 | 2.70 | 2.87 | 3.19 |
| $stdev(\Omega_{NaO_3})$ | 0.730 | 0.778 | 0.767 | 0.884 |
| $\langle r_M \rangle$ (Å) | 0.393 | 0.380 | 0.420 | 0.428 |
| $V$ (Å$^3$) | 2130 | 2070 | 2140 | 2200 |

(to be continued)

| Descriptor | $Na_3Zr_{1.25}Ti_{0.75}Si_2PO_{12}$ | $Na_3Zr_{1.75}Ti_{0.25}Si_2PO_{12}$ | $Na_3Zr_{1.75}Sn_{0.25}Si_2PO_{12}$ | Mean | Standard deviation |
|---|---|---|---|---|---|
| $\langle C_P \rangle$ | -0.583 | -0.583 | -0.583 | -0.458 | 0.209 |
| $\langle \chi_M \rangle$ | 1.76 | 1.74 | 1.76 | 1.82 | 0.123 |
| $d_0$ (Å) | 2.44 | 2.49 | 2.49 | 2.53 | 0.0824 |
| $d_1$ (Å) | 1.67 | 1.63 | 1.67 | 1.66 | 0.0882 |
| $V_p$ (Å$^3$) | 7.54 | 9.61 | 9.93 | 11.0 | 8.00 |
| $\rho_{Na}$ (Å$^{-3}$) | 0.0112 | 0.0111 | 0.0110 | 0.00922 | 0.00364 |
| $n$ | 3 | 3 | 3 | 2.49 | 1.04 |
| $\langle d_{Na-Na} \rangle$ (Å) | 3.57 | 3.54 | 3.55 | 3.64 | 0.164 |
| $med(d_{Na-Na})$ (Å) | 3.52 | 3.52 | 3.53 | 3.63 | 0.175 |
| $\langle d_{Na-O} \rangle$ (Å) | 2.56 | 2.57 | 2.57 | 2.56 | 0.0312 |
| $\langle n_{Na-O} \rangle$ | 5.39 | 5.57 | 5.52 | 5.29 | 0.245 |
| $\langle V_{NaO_x} \rangle$ (Å$^3$) | 18.1 | 19.1 | 18.4 | 17.8 | 1.13 |
| $\langle \min(\Omega_{NaO_3}) \rangle$ | 0.619 | 0.658 | 0.611 | 1.42 | 2.86 |
| $\langle \Omega_{NaO_3} \rangle$ | 1.43 | 1.40 | 1.42 | 1.47 | 0.0629 |
| $\langle \max(\Omega_{NaO_3}) \rangle$ | 2.99 | 2.79 | 2.86 | 2.81 | 0.428 |
| $stdev(\Omega_{NaO_3})$ | 0.840 | 0.783 | 0.860 | 0.786 | 0.132 |
| $\langle r_M \rangle$ (Å) | 0.409 | 0.420 | 0.425 | 0.407 | 0.0429 |
| $V$ (Å$^3$) | 2130 | 2170 | 2190 | 2140 | 133 |

| Descriptor | $Na_{2.75}Zr_{1.75}Nb_{0.25}Si_2PO_{12}$ |
|---|---|
| $\langle C_P \rangle$ | -0.556 |
| $\langle \chi_M \rangle$ | 1.74 |
| $d_0$ (Å) | 2.54 |
| $d_1$ (Å) | 1.70 |
| $V_p$ (Å$^3$) | 10.3 |
| $\rho_{Na}$ (Å$^{-3}$) | 0.0101 |
| $n$ | 2.75 |
| $\langle d_{Na-Na} \rangle$ (Å) | 3.56 |
| $med(d_{Na-Na})$ (Å) | 3.54 |
| $\langle d_{Na-O} \rangle$ (Å) | 2.57 |
| $\langle n_{Na-O} \rangle$ | 5.30 |
| $\langle V_{NaO_x} \rangle$ (Å$^3$) | 17.8 |
| $\langle \min(\Omega_{NaO_3}) \rangle$ | 0.678 |
| $\langle \Omega_{NaO_3} \rangle$ | 1.49 |
| $\langle \max(\Omega_{NaO_3}) \rangle$ | 3.08 |
| $stdev(\Omega_{NaO_3})$ | 0.830 |
| $\langle r_M \rangle$ (Å) | 0.422 |
| $V$ (Å$^3$) | 2180 |

**Supplementary Note 2** Maximum likelihood estimation and pseudo-goodnees-of-fit $R^2_{\text{pseudo}}$ for the exhaustive search of the possible multivariate beta regression models

In beta regression modelling, it is assumed that each $\eta_j$ (score of $D_{\text{Na,300K}}$ for datum $j$: $j = 1, \cdots n_{\text{data}}$) follows the beta distribution: $\eta_j \sim \text{B}(\eta_j; \bar{\eta}_j, \varphi)$:

$$\text{B}(\eta_j; \bar{\eta}_j, \varphi) = \frac{\int_0^\infty t^{\varphi-1} e^{-t} dt}{\int_0^\infty t^{\bar{\eta}_j \varphi - 1} e^{-t} dt \int_0^\infty t^{[(1-\bar{\eta}_j)\varphi]-1} e^{-t} dt} \eta_j^{\bar{\eta}_j \varphi - 1} (1 - \eta_j)^{(1-\bar{\eta}_j)\varphi - 1}, \quad (\text{S1})$$

where $\bar{\eta}_j$ and $\varphi$ denote the most-fit value $\bar{\eta}$ for $\eta_j$ (given as the mean for B) and the common precision for B, respectively.[4, 5] Then, by executing maximum likelihood estimation with parameters $c(x_i)$, $c$, and $\varphi$ towards the minimization of the sum $\mathcal{L}$ of the logarithm-scaled beta densities (i.e., the log-likelihood function proposed by Ferrari and Cribari-Neto)[5] given by

$$\mathcal{L} = \sum_{j=1}^{n_{\text{data}}} \log_e [\text{B}(\eta_j; \bar{\eta}_j, \varphi)], \quad (\text{S2})$$

the most-fitting $h$ (consequently, $\bar{\eta}$) would be found from the input $\eta$ and $z[x_i]$.

We set $df = 1$ and 2, with which we exhaustively examined 153 $[= \sum_{df=1,2} \binom{n(x_i)}{df}$; see the number $n(x_i)$ of features (descriptor candidates) $x_i$ is 17] multivariate regression models and picked the two models with the highest pseudo-goodnees-of-fit values $R^2_{\text{pseudo}}$:[5]

$$R^2_{\text{pseudo}} = R^2 [h(\bar{\eta}_j), h(\eta_j)], \quad (\text{S3})$$

that is, the $R^2$ value between $h(\bar{\eta}_j)$ and $h(\eta_j)$: $h(\bar{\eta}_j) = \log_e \frac{\bar{\eta}_j}{1-\bar{\eta}_j}$ and $h(\eta_j) = \log_e \frac{\eta_j}{1-\eta_j}$, which is the inverse function of Eq. (11) (i.e., the logit function). In addition to $R^2_{\text{pseudo}}$, we added criteria that $p$-values for $c(x_i)$, $c$, and $\varphi$ should be less than .05. Also, to examine any multicollinearity issue between $x_i$, we also computed the Pearson correlation coefficient between the two taken descriptors and $x_1$ and $x_2$:

$$r(x_1, x_2) = \frac{\sum_{j=1}^{n_{\text{data}}}(x_1-\overline{x_1})(x_2-\overline{x_2})}{\sqrt{\sum_{j=1}^{n_{\text{data}}}(x_1-\overline{x_1})^2}\sqrt{\sum_{j=1}^{n_{\text{data}}}(x_2-\overline{x_2})^2}}, \quad (S4)$$

where $\overline{x_i}$ is the mean value of $x_1$.

**Supplementary Note 3** Descriptors for $D_{\text{Na,300K}}$ in NASICONs and Na-ion sulfides

In our previous study on Na-ion sulfides, where tetrahedral units $MS_4$ are separated, we conducted a statistical analysis and found that structural distortions in $\text{NaS}_x$, specifically towards wide faces [indicated by large $\langle \max(\Omega_{\text{NaS}_3}) \rangle$], along with shallow electrostatic potential (by small $\langle C_P \rangle$), play a crucial role in increasing $D_{\text{Na,300K}}$.[6] We further confirmed the positive influence of such structural distortions on $D_{\text{Na,300K}}$ in NASICON samples through another beta regression model. The multivariate beta regression model with the highest $R^2_{\text{pseudo}}$ given two descriptors allowed [one of which was set to $\langle \max(\Omega_{\text{NaO}_3}) \rangle$] was identified as

$$h = 0.631\, z[\langle V_{\text{NaO}_x} \rangle] + 0.585\, z[\langle \max(\Omega_{\text{NaO}_3}) \rangle] + 0.488 \qquad \text{(S5)}$$

with $\varphi = 3.23 : R^2_{\text{pseudo}} = .454$. The Pearson correlation coefficient between $\langle V_{\text{NaO}_x} \rangle$ and $\langle \max(\Omega_{\text{NaO}_3}) \rangle$ was given as $r(\langle V_{\text{NaO}_x} \rangle, \langle V_{\text{NaO}_x} \rangle) = .00402$, indicating the absence of the significant multicollinearity issue within Eq. (S5). Given the higher $R^2_{\text{pseudo}}$ value for Eq. (2), namely $R^2_{\text{pseudo}} = 0.797$, compared to Eq. (S5), it is suggested that modulating $d_1$ and $\langle d_{\text{Na-Na}} \rangle$, rather than $\langle \max(\Omega_{\text{NaO}_3}) \rangle$, is more favorable to achieve higher $D_{\text{Na,300K}}$ for NASICONs. This implies that, in the case of NASICON structures, enhancing the already-established diffusion pathways indicated by the underlying "skeleton" structure (represented by $d_1$) and optimizing the proximity of Na-ions along these pathways ($\langle d_{\text{Na-Na}} \rangle$) are advantageous. It is suggested that the focus should not solely be on designing local structures associated with polyhedra $\text{NaO}_x$, but rather on manipulating the overall framework to facilitate efficient ion mobility, especially for NASICONs. Meanwhile, it is worth noting that NASICONs taking high-valence ions, such as As or P, already exhibit small $\langle C_\mathrm{P} \rangle$ values within the range of [-0.722, -0.184], whereas Na-ion sulfides display larger and more widely distributed $\langle C_\mathrm{P} \rangle$ values within the range of [-6, -2.5].[6]

**Supplementary Note 4** Linear regressions for the primary descriptors against the secondary ones

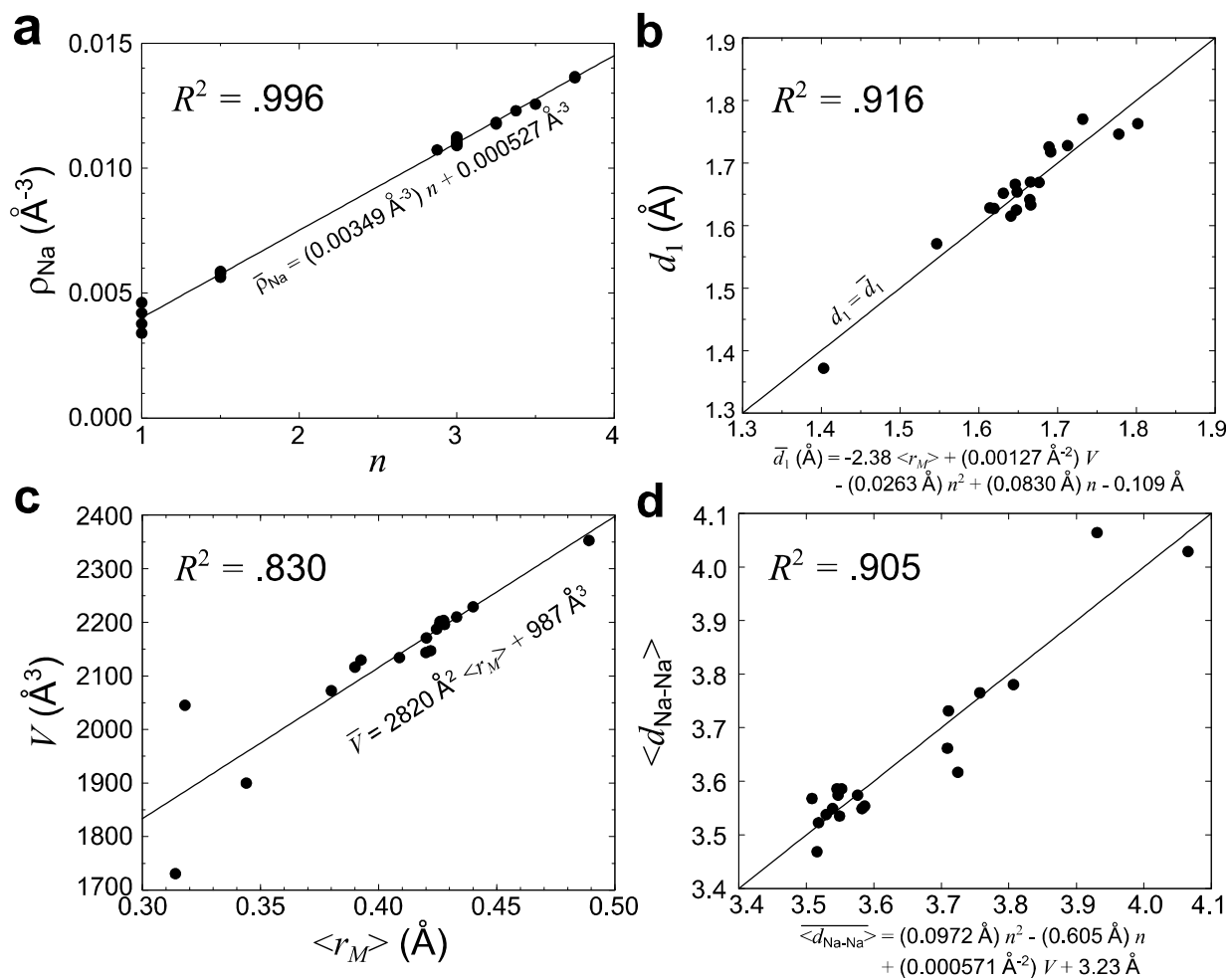

**Supplementary Fig. 4** Linear regression models for $\rho_{Na}$, $d_1$, $V$, and $\langle d_{Na-N} \rangle$ given in Eqs. (S6)—(S9). The dots are estimated from the unit cell samples of which geometry was optimized by density functional theory (DFT) calculations, and the lines show the regressed (simulated) results.

Our goal is to develop models wherein the Na-ion density $\rho_{Na}$, the bottleneck width $d_1$, and the average Na-Na bond length $\langle d_{Na-} \rangle$ eventually are connected to the Na-ion content $n$ and the average $\langle r_M \rangle$ of ionic radii $r_M$ for metal ions (excluding Na-ions),[3] with $R^2 > 0.9$ if attainable, to keep the internal consistency of the whole modelling. We briefly note that the potential multicollinearity problem is not of significant concern, as far as these equations play a crucial role in predicting $\sigma_{Na,300K}$ eventually. The results were given below;

$$\rho_{Na} = (0.00349 \text{ Å}^{-3})n + 0.000527 \text{ Å}^{-3}, \qquad (S6)$$

$$d_1 = -2.38 \langle r_M \rangle + (0.00127 \text{ Å}^{-2})V - (0.0263 \text{ Å})n^2 + (0.0830 \text{ Å})n - 0.109 \text{ Å}, \qquad (S7)$$

$$V = (2820 \text{ Å}^2)\langle r_M \rangle + 987 \text{ Å}^3, \qquad (S8)$$

and

$$\langle d_{Na-Na} \rangle = (0.0972 \text{ Å})n^2 - (0.605 \text{ Å})n + (0.000571 \text{ Å}^{-2})V + 3.23 \text{ Å}, \qquad (S9)$$

where $V$ denotes the volume for a unit cell of each sample. $\rho_{Na}$ will increase with $n$. $R^2$ for Eqs. (S6)—(S9) were given as 0.996, 0.916, 0.830, and 0.905, respectively. In Supplementary Fig. 4a and Eq. (S6), $n$ suffices the regression modelling for $\rho_{Na}$ without $V$. In Supplementary Fig. 4c and Eq. (S8), $R^2$ rises to 0.943 subsequent to the removal of an outlier, the case of $Na_{1.5}Zr_{1.5}Al_{0.5}P_3O_{12}$ characterized by the compressed $V = 1880 \text{ Å}^3$. It is noteworthy that, $R^2 > 0.9$ for $V$ necessitated the incorporation of interactive terms $(n\langle r_M \rangle)^k$ with $k = 1$ and 2.

$d_1$ would be maximized by taking a small $\langle r_M \rangle$ in $V$ (expanded by, on the contrary, large $\langle r_M \rangle$) and optimizing the Na-ion content $n$ around 1.58. $\langle d_{Na-N} \rangle$ would be minimized by taking a small $V$ and optimizing $n$ around 3.11. Hence, the optimizing values $n$ for $d_1$ and $\langle d_{Na-Na} \rangle$ are opposite, and other related factors such as $\langle r_\text{M} \rangle$ and $V$ (that is regressed against $\langle r_\text{M} \rangle$ again) should be considered as well to model the room-temperature Na-ion conductivity $\sigma_\text{Na,300K}$ correctly.

**Supplementary Note 5** Single-$T$ "long-time" diagnosis for $Na_{2.75}Zr_{1.75}Nb_{0.25}Si_2PO_{12}$

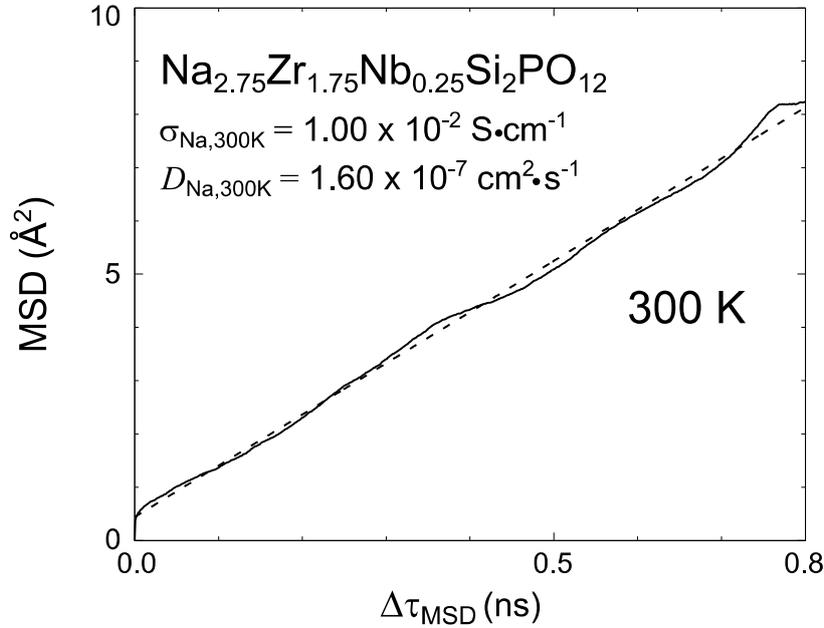

**Supplementary Fig. 5** Mean squared displacement curve against sampled time intervals $\Delta\tau_{MSD}$ given by the single-$T$ "long-time" diagnosis (with $\Delta\tau$ = 1 fs and $\tau$ = 1 ns at $T$ = 300 K) for $Na_{2.75}Zr_{1.75}Nb_{0.25}Si_2PO_{12}$. The black dashed line represents regression against sampled time intervals $\Delta\tau_{MSD}$.

We conducted the single-$T$ "long-time" diagnosis for $Na_{2.75}Zr_{1.75}Nb_{0.25}Si_2PO_{12}$. The obtained results were highly remarkable in an unexpected way, as evidenced by $\sigma_{Na,300K}$ = $1.00 \times 10^{-2}$ S·cm$^{-1}$ and $D_{Na,300K}$ = $1.60 \times 10^{-7}$ cm$^2$·s$^{-1}$. In Supplementary Fig. 5, we depict the nearly linear MSD curves $R^2_{MSD}$ = 0.997. We postulate that the disparities observed in the values of $\sigma_{Na,300K}$ between the two approaches (that is, the single-$T$ "long-time" diagnosis and the multi-

$T$ DFT-MD simulations) can be attributed to a technical issue, specifically, the contraction of unit cell volumes $V$ during the $NpT$ pre-treatment stage for the latter: 4.66 %. While $V$ was fixed to the geometry-optimized one in the single-$T$ "long-time" diagnosis, the multi-$T$ DFT-MD simulations employed the unit cell whose $V$ was thermally equilibrated during the $NpT$ pre-treatment to include the high-$T$ effect for $E_a$ more accurately. On the contrary, the different choice of $\boldsymbol{k}$-grids and the kinetic energy cutoff between the geometry optimization and the $NpT$ pre-treatment at low $T$ may have elicited the $V$ contraction and sensitive steric hindrance effects for the self-diffusion of Na-ions. Given the prevalent overestimation problem associated with lattice constants in the generalized gradient approximation scheme,[7, 8] the true $\sigma_{\text{Na,300K}}$ likely resides between the two $\sigma_{\text{Na,300K}}$ (that is, still in the order of $10^{-3}$ S·cm$^{-1}$), which may be addressed in future study with experimental supports. Nonetheless, in both approaches, our findings assert that the structurally-stable $\text{Na}_{2.75}\text{Zr}_{1.75}\text{Nb}_{0.25}\text{Si}_2\text{PO}_{12}$, which hitherto remain unexplored and validate the predictability of our model $\sigma_{\text{Na,300K,sim}}$, hold substantial promise for future investigation.

**Supplementary Table 3** Values for $n$, $\langle r_M \rangle$, and the experimental values for $\sigma_{Na,300K}$, which were referred to by Eqs. (7) and (8). $\sigma_{Na,300K}$ were interpolated or measured at $T = 300$ K[9-19] or extrapolated to $T = 300$ K.[20-34]

| Chemical formula | | | | | | | | | | | | | | | $n$ | $\langle r_M \rangle$ | experimental $\sigma_{Na,300K}$ (S·cm$^{-1}$) |
|---|---|---|---|---|---|---|---|---|---|---|---|---|---|---|---|---|---|
| Na | 1   | Zr | 2    | Y  | 1    |    |     | P | 3   | O | 12    | 1   | 0.475 | $1.60 \times 10^{-7}$ [9] |
| Na | 1.5 | Zr | 1.5  | Y  | 0.5  |    |     | P | 3   | O | 12    | 1.5 | 0.408 | $2.73 \times 10^{-6}$ [9] |
| Na | 2   | Zr | 1    | Y  | 1    |    |     | P | 3   | O | 12    | 2   | 0.426 | $8.55 \times 10^{-6}$ [9] |
| Na | 2.5 | Zr | 0.5  | Y  | 1.5  |    |     | P | 3   | O | 12    | 2.5 | 0.444 | $9.09 \times 10^{-7}$ [9] |
| Na | 3   | Zr | 2    |    |      | Si | 2   | P | 1   | O | 12    | 3   | 0.426 | $4.00 \times 10^{-3}$ [10] |
| Na | 3.1 | Zr | 1.55 |    |      | Si | 2.3 | P | 0.7 | O | 11    | 3.1 | 0.403 | $5.00 \times 10^{-4}$ [10] |
| Na | 3.1 | Zr | 2    | Al | 0.1  | Si | 1.9 | P | 1   | O | 12    | 3.1 | 0.432 | $2.00 \times 10^{-4}$ [10] |
| Na | 3.2 | Zr | 2    | Al | 0.2  | Si | 1.8 | P | 1   | O | 12    | 3.2 | 0.437 | $2.00 \times 10^{-4}$ [10] |
| Na | 3.3 | Zr | 1.55 | Al | 0.2  | Si | 2.1 | P | 0.7 | O | 11    | 3.3 | 0.415 | $4.00 \times 10^{-4}$ [10] |
| Na | 3.4 | Zr | 1.55 | Al | 0.3  | Si | 2   | P | 0.7 | O | 11    | 3.4 | 0.421 | $1.00 \times 10^{-4}$ [10] |
| Na | 2.4 |    |      | Hf | 2    | Si | 1.4 | P | 1.6 | O | 12    | 2.4 | 0.411 | $7.30 \times 10^{-4}$ [11] |
| Na | 2.6 |    |      | Hf | 2    | Si | 1.6 | P | 1.4 | O | 12    | 2.6 | 0.415 | $5.90 \times 10^{-4}$ [11] |
| Na | 2.8 |    |      | Hf | 2    | Si | 1.8 | P | 1.2 | O | 12    | 2.8 | 0.418 | $6.90 \times 10^{-4}$ [11] |
| Na | 3   |    |      | Hf | 2    | Si | 2   | P | 1   | O | 12    | 3   | 0.422 | $1.10 \times 10^{-3}$ [11] |
| Na | 3.2 |    |      | Hf | 2    | Si | 2.2 | P | 0.8 | O | 12    | 3.2 | 0.426 | $2.30 \times 10^{-3}$ [11] |
| Na | 3.4 |    |      | Hf | 2    | Si | 2.4 | P | 0.6 | O | 12    | 3.4 | 0.429 | $1.40 \times 10^{-3}$ [11] |
| Na | 3.6 |    |      | Hf | 2    | Si | 2.6 | P | 0.4 | O | 12    | 3.6 | 0.433 | $1.20 \times 10^{-3}$ [11] |
| Na | 3.8 |    |      | Hf | 2    | Si | 2.8 | P | 0.2 | O | 12    | 3.8 | 0.436 | $3.20 \times 10^{-4}$ [11] |
| Na | 3   | Zr | 1.88 | Si | 2    | Y  | 0.12 | P | 1  | O | 11.94 | 3   | 0.430 | $2.50 \times 10^{-3}$ [12] |
| Na | 3.4 |    |      | Sc | 2    | Si | 0.4 | P | 2.6 | O | 12    | 3.4 | 0.407 | $6.90 \times 10^{-4}$ [13] |
| Na | 3.1 | Zr | 1.95 | Mg | 0.05 | Si | 2   | P | 1   | O | 12    | 3.1 | 0.426 | $3.50 \times 10^{-3}$ [14] |
| Na | 3   | Zr | 2    |    |      | Si | 2   | P | 1   | O | 12    | 3   | 0.426 | $1.10 \times 10^{-4}$ [15] |
| Na | 3   | Zr | 1.9  | Yb | 0.1  | Si | 2   | P | 1   | O | 12    | 3   | 0.429 | $1.70 \times 10^{-4}$ [15] |
| Na | 3   | Zr | 1.9  | Gd | 0.1  | Si | 2   | P | 1   | O | 12    | 3   | 0.430 | $6.00 \times 10^{-4}$ [15] |
| Na | 3   | Zr | 1.9  | Ce | 0.1  | Si | 2   | P | 1   | O | 12    | 3   | 0.429 | $9.00 \times 10^{-4}$ [15] |
| Na | 3   | Zr | 2    |    |      | Si | 2   | P | 1   | O | 12    | 3   | 0.426 | $2.00 \times 10^{-3}$ [16] |
| Na | 3.2 | Zr | 1.8  | Sc | 0.2  | Si | 2   | P | 1   | O | 12    | 3.2 | 0.427 | $5.30 \times 10^{-3}$ [16] |
| Na | 3.4 | Zr | 1.6  | Sc | 0.4  | Si | 2   | P | 1   | O | 12    | 3.4 | 0.428 | $6.20 \times 10^{-3}$ [16] |
| Na | 3.6 | Zr | 1.4  | Sc | 0.6  | Si | 2   | P | 1   | O | 12    | 3.6 | 0.429 | $5.10 \times 10^{-3}$ [16] |

| | | | | | | | | | | | | |
|---|---|---|---|---|---|---|---|---|---|---|---|---|
| Na | 3    | Zr | 2    |    |      | Si | 2   | P | 1   | O | 12 | 3    | 0.426 | $3.80 \times 10^{-4}$ [17] |
| Na | 3.1  | Zr | 1.95 | Ca | 0.05 | Si | 2   | P | 1   | O | 12 | 3.1  | 0.429 | $7.58 \times 10^{-4}$ [17] |
| Na | 3.2  | Zr | 1.9  | Ca | 0.1  | Si | 2   | P | 1   | O | 12 | 3.2  | 0.432 | $1.67 \times 10^{-3}$ [17] |
| Na | 3.3  | Zr | 1.85 | Ca | 0.15 | Si | 2   | P | 1   | O | 12 | 3.3  | 0.434 | $1.33 \times 10^{-3}$ [17] |
| Na | 3.4  | Zr | 1.8  | Ca | 0.2  | Si | 2   | P | 1   | O | 12 | 3.4  | 0.437 | $6.41 \times 10^{-4}$ [17] |
| Na | 3.5  | Zr | 1.75 | Ca | 0.25 | Si | 2   | P | 1   | O | 12 | 3.5  | 0.440 | $9.95 \times 10^{-4}$ [17] |
| Na | 3.3  | Zr | 2    |    |      | Si | 2.3 | P | 0.7 | O | 12 | 3.3  | 0.431 | $2.14 \times 10^{-3}$ [18] |
| Na | 3.35 | Zr | 1.95 | Nb | 0.05 | Si | 2.4 | P | 0.6 | O | 12 | 3.35 | 0.432 | $4.06 \times 10^{-3}$ [18] |
| Na | 3.3  | Zr | 1.9  | Nb | 0.1  | Si | 2.4 | P | 0.6 | O | 12 | 3.3  | 0.432 | $5.51 \times 10^{-3}$ [18] |
| Na | 3.25 | Zr | 1.85 | Nb | 0.15 | Si | 2.4 | P | 0.6 | O | 12 | 3.25 | 0.431 | $4.44 \times 10^{-3}$ [18] |
| Na | 3.2  | Zr | 1.8  | Nb | 0.2  | Si | 2.4 | P | 0.6 | O | 12 | 3.2  | 0.430 | $2.94 \times 10^{-3}$ [18] |
| Na | 3.1  | Zr | 1.7  | Nb | 0.3  | Si | 2.4 | P | 0.6 | O | 12 | 3.1  | 0.428 | $2.69 \times 10^{-3}$ [18] |
| Na | 3    | Zr | 1.6  | Nb | 0.4  | Si | 2.4 | P | 0.6 | O | 12 | 3    | 0.427 | $1.39 \times 10^{-3}$ [18] |
| Na | 3.36 | Zr | 1.96 | Nb | 0.04 | Si | 2.4 | P | 0.6 | O | 12 | 3.36 | 0.433 | $1.61 \times 10^{-3}$ [19] |
| Na | 2.96 | Zr | 1.96 | Nb | 0.04 | Si | 2   | P | 1   | O | 12 | 2.96 | 0.425 | $4.15 \times 10^{-4}$ [19] |
| Na | 3.4  | Zr | 2    |    |      | Si | 2.4 | P | 0.6 | O | 12 | 3.4  | 0.433 | $3.65 \times 10^{-4}$ [19] |
| Na | 3    | Zr | 2    |    |      | Si | 2   | P | 1   | O | 12 | 3    | 0.426 | $4.21 \times 10^{-4}$ [19] |
| Na | 3.08 | Zr | 1.96 | Mg | 0.04 | Si | 2   | P | 1   | O | 12 | 3.08 | 0.426 | $1.10 \times 10^{-3}$ [20] |
| Na | 3.2  | Zr | 1.9  | Mg | 0.1  | Si | 2   | P | 1   | O | 12 | 3.2  | 0.426 | $4.97 \times 10^{-4}$ [20] |
| Na | 3.6  | Zr | 1.7  | Mg | 0.3  | Si | 2   | P | 1   | O | 12 | 3.6  | 0.426 | $4.91 \times 10^{-5}$ [20] |
| Na | 3.08 | Zr | 1.96 | Zn | 0.04 | Si | 2   | P | 1   | O | 12 | 3.08 | 0.426 | $4.79 \times 10^{-5}$ [20] |
| Na | 3.2  | Zr | 1.9  | Zn | 0.1  | Si | 2   | P | 1   | O | 12 | 3.2  | 0.426 | $1.97 \times 10^{-4}$ [20] |
| Na | 3.4  | Zr | 1.8  | Zn | 0.2  | Si | 2   | P | 1   | O | 12 | 3.4  | 0.427 | $5.81 \times 10^{-4}$ [20] |
| Na | 3.6  | Zr | 1.7  | Zn | 0.3  | Si | 2   | P | 1   | O | 12 | 3.6  | 0.427 | $3.14 \times 10^{-5}$ [20] |
| Na | 3.04 | Zr | 1.96 | Y  | 0.04 | Si | 2   | P | 1   | O | 12 | 3.04 | 0.427 | $3.54 \times 10^{-4}$ [20] |
| Na | 3.1  | Zr | 1.9  | Y  | 0.1  | Si | 2   | P | 1   | O | 12 | 3.1  | 0.430 | $2.08 \times 10^{-3}$ [20] |
| Na | 3.3  | Zr | 1.7  | Y  | 0.3  | Si | 2   | P | 1   | O | 12 | 3.3  | 0.437 | $1.16 \times 10^{-4}$ [20] |
| Na | 3    | Zr | 1.9  | Ti | 0.1  | Si | 2   | P | 1   | O | 12 | 3    | 0.424 | $6.67 \times 10^{-5}$ [20] |
| Na | 3    | Zr | 1.8  | Ti | 0.2  | Si | 2   | P | 1   | O | 12 | 3    | 0.421 | $1.25 \times 10^{-4}$ [20] |
| Na | 3    | Zr | 1.7  | Ti | 0.3  | Si | 2   | P | 1   | O | 12 | 3    | 0.419 | $9.68 \times 10^{-5}$ [20] |
| Na | 2.96 | Zr | 1.96 | Nb | 0.04 | Si | 2   | P | 1   | O | 12 | 2.96 | 0.425 | $3.08 \times 10^{-3}$ [20] |
| Na | 2.9  | Zr | 1.9  | Nb | 0.1  | Si | 2   | P | 1   | O | 12 | 2.9  | 0.424 | $5.31 \times 10^{-4}$ [20] |
| Na | 2.96 | Zr | 1.96 | Ta | 0.04 | Si | 2   | P | 1   | O | 12 | 2.96 | 0.425 | $1.97 \times 10^{-3}$ [20] |
| Na | 2.9  | Zr | 1.9  | Ta | 0.1  | Si | 2   | P | 1   | O | 12 | 2.9  | 0.424 | $6.29 \times 10^{-4}$ [20] |
| Na | 2.96 | Zr | 1.96 | V  | 0.04 | Si | 2   | P | 1   | O | 12 | 2.96 | 0.425 | $2.34 \times 10^{-3}$ [20] |

| | | | | | | | | | | | | | | | |
|---|---|---|---|---|---|---|---|---|---|---|---|---|---|---|---|
| Na | 2.9 | Zr | 1.9 | V | 0.1 | Si | 2 | P | 1 | | | O | 12 | 2.9 | 0.422 | $7.62 \times 10^{-4}$ [20] |
| Na | 1 | Zr | 2 | | | | | P | 3 | | | O | 12 | 1 | 0.390 | $4.61 \times 10^{-9}$ [21] |
| Na | 1.5 | Zr | 1.5 | In | 0.5 | | | P | 3 | | | O | 12 | 1.5 | 0.398 | $1.29 \times 10^{-7}$ [21] |
| Na | 1.8 | Zr | 1.2 | In | 0.8 | | | P | 3 | | | O | 12 | 1.8 | 0.403 | $3.13 \times 10^{-7}$ [21] |
| Na | 2 | Zr | 1 | In | 1 | | | P | 3 | | | O | 12 | 2 | 0.406 | $6.60 \times 10^{-7}$ [21] |
| Na | 2.2 | Zr | 0.8 | In | 1.2 | | | P | 3 | | | O | 12 | 2.2 | 0.409 | $1.87 \times 10^{-6}$ [21] |
| Na | 2.5 | Zr | 0.5 | In | 1.5 | | | P | 3 | | | O | 12 | 2.5 | 0.414 | $1.77 \times 10^{-6}$ [21] |
| Na | 2.75 | Zr | 0.25 | In | 1.75 | | | P | 3 | | | O | 12 | 2.75 | 0.418 | $2.42 \times 10^{-6}$ [21] |
| Na | 1.5 | Zr | 1.5 | Yb | 0.5 | | | P | 3 | | | O | 12 | 1.5 | 0.405 | $1.30 \times 10^{-6}$ [21] |
| Na | 1.8 | Zr | 1.2 | Yb | 0.8 | | | P | 3 | | | O | 12 | 1.8 | 0.414 | $4.02 \times 10^{-6}$ [21] |
| Na | 2 | Zr | 1 | Yb | 1 | | | P | 3 | | | O | 12 | 2 | 0.420 | $6.47 \times 10^{-6}$ [21] |
| Na | 2.2 | Zr | 0.8 | Yb | 1.2 | | | P | 3 | | | O | 12 | 2.2 | 0.426 | $5.23 \times 10^{-6}$ [21] |
| Na | 2.3 | Zr | 0.7 | Yb | 1.3 | | | P | 3 | | | O | 12 | 2.3 | 0.428 | $6.77 \times 10^{-6}$ [21] |
| Na | 2.5 | Zr | 0.5 | Yb | 1.5 | | | P | 3 | | | O | 12 | 2.5 | 0.434 | $2.91 \times 10^{-6}$ [21] |
| Na | 2.6 | Zr | 0.4 | Yb | 1.6 | | | P | 3 | | | O | 12 | 2.6 | 0.437 | $6.43 \times 10^{-7}$ [21] |
| Na | 2.8 | Zr | 0.2 | Yb | 1.8 | | | P | 3 | | | O | 12 | 2.8 | 0.443 | $3.17 \times 10^{-7}$ [21] |
| Na | 2.9 | Zr | 0.1 | Yb | 1.9 | | | P | 3 | | | O | 12 | 2.9 | 0.446 | $2.57 \times 10^{-8}$ [21] |
| Na | 1.5 | Zr | 1.5 | Cr | 0.5 | | | P | 3 | | | O | 12 | 1.5 | 0.380 | $4.81 \times 10^{-7}$ [21] |
| Na | 2 | Zr | 1 | Cr | 1 | | | P | 3 | | | O | 12 | 2 | 0.369 | $2.07 \times 10^{-6}$ [21] |
| Na | 2.3 | Zr | 0.7 | Cr | 1.3 | | | P | 3 | | | O | 12 | 2.3 | 0.363 | $5.91 \times 10^{-6}$ [21] |
| Na | 2.5 | Zr | 0.5 | Cr | 1.5 | | | P | 3 | | | O | 12 | 2.5 | 0.359 | $9.92 \times 10^{-6}$ [21] |
| Na | 2.7 | Zr | 0.3 | Cr | 1.7 | | | P | 3 | | | O | 12 | 2.7 | 0.354 | $6.04 \times 10^{-6}$ [21] |
| Na | 2.95 | Zr | 0.05 | Cr | 1.95 | | | P | 3 | | | O | 12 | 2.95 | 0.349 | $2.40 \times 10^{-6}$ [21] |
| Na | 3 | Cr | 2 | | | | | P | 3 | | | O | 12 | 3 | 0.348 | $3.63 \times 10^{-9}$ [21] |
| Na | 3 | Zr | 2 | | | Si | 2 | P | 0.6 | As | 0.4 | O | 12 | 3 | 0.439 | $6.66 \times 10^{-4}$ [22] |
| Na | 3 | Zr | 1.6 | Ti | 0.4 | | | P | 1 | | | O | 12 | 3 | 0.521 | $4.12 \times 10^{-4}$ [22] |
| Na | 3 | Zr | 2 | Ge | 0.8 | Si | 1.2 | P | 0.6 | As | 0.4 | O | 12 | 3 | 0.482 | $6.44 \times 10^{-4}$ [22] |
| Na | 3 | Zr | 1.6 | Th | 0.4 | Si | 2 | P | 1 | | | O | 12 | 3 | 0.444 | $9.21 \times 10^{-4}$ [22] |
| Na | 3 | Fe | 2 | | | | | P | 3 | | | O | 12 | 3 | 0.360 | $1.17 \times 10^{-8}$ [23] |
| Na | 3 | Cr | 2 | | | | | P | 3 | | | O | 12 | 3 | 0.348 | $1.76 \times 10^{-8}$ [23] |
| Na | 1 | Zr | 2 | | | | | | | As | 3 | O | 12 | 1 | 0.489 | $4.11 \times 10^{-7}$ [24] |
| Na | 1.15 | Zr | 0.85 | Yb | 0.15 | | | | | As | 3 | O | 12 | 1.15 | 0.437 | $7.85 \times 10^{-7}$ [24] |
| Na | 1.25 | Zr | 0.75 | Yb | 0.25 | | | | | As | 3 | O | 12 | 1.25 | 0.441 | $1.10 \times 10^{-6}$ [24] |
| Na | 1.35 | Zr | 0.65 | Yb | 0.35 | | | | | As | 3 | O | 12 | 1.35 | 0.444 | $1.44 \times 10^{-6}$ [24] |
| Na | 1.55 | Zr | 0.55 | Yb | 0.45 | | | | | As | 3 | O | 12 | 1.55 | 0.448 | $1.83 \times 10^{-6}$ [24] |
| Na | 3 | Zr | 0.5 | Sc | 1.5 | Si | 0.5 | P | 1.5 | | | O | 12 | 3 | 0.466 | $7.42 \times 10^{-5}$ [25] |

| | | | | | | | | | | | | | |
|---|---|---|---|---|---|---|---|---|---|---|---|---|---|
| Na | 2.7 | Zr | 1.8 | Sc | 0.2 | Si | 1.5 | P | 1.5 | O | 12 | 2.7 | 0.418 | $2.46 \times 10^{-4}$ [25] |
| Na | 3 | Zr | 1.8 | Sc | 0.2 | Si | 1.8 | P | 1.2 | O | 12 | 3 | 0.423 | $3.40 \times 10^{-4}$ [25] |
| Na | 1 | Zr | 2 | | | | | P | 3 | O | 12 | 1 | 0.390 | $3.98 \times 10^{-9}$ [26] |
| Na | 1.3 | Zr | 1.85 | Mg | 0.15 | | | P | 3 | O | 12 | 1.3 | 0.390 | $1.03 \times 10^{-8}$ [26] |
| Na | 1.6 | Zr | 1.7 | Mg | 0.3 | | | P | 3 | O | 12 | 1.6 | 0.390 | $2.23 \times 10^{-8}$ [26] |
| Na | 1.8 | Zr | 1.6 | Mg | 0.4 | | | P | 3 | O | 12 | 1.8 | 0.390 | $2.28 \times 10^{-7}$ [26] |
| Na | 2 | Zr | 1.5 | Mg | 0.5 | | | P | 3 | O | 12 | 2 | 0.390 | $3.72 \times 10^{-7}$ [26] |
| Na | 2.5 | Zr | 1.25 | Mg | 0.75 | | | P | 3 | O | 12 | 2.5 | 0.390 | $7.94 \times 10^{-7}$ [26] |
| Na | 3 | Zr | 1 | Mg | 1 | | | P | 3 | O | 12 | 3 | 0.390 | $1.21 \times 10^{-6}$ [26] |
| Na | 3.2 | Zr | 2 | Si | 2.2 | | | P | 0.8 | O | 12 | 3.2 | 0.430 | $5.01 \times 10^{-4}$ [27] |
| Na | 3.28 | Zr | 1.96 | Mg | 0.04 | Si | 2.2 | P | 0.8 | O | 12 | 3.28 | 0.430 | $5.01 \times 10^{-4}$ [27] |
| Na | 3.36 | Zr | 1.92 | Mg | 0.08 | Si | 2.2 | P | 0.8 | O | 12 | 3.36 | 0.430 | $5.01 \times 10^{-4}$ [27] |
| Na | 3.52 | Zr | 1.84 | Mg | 0.16 | Si | 2.2 | P | 0.8 | O | 12 | 3.52 | 0.430 | $2.58 \times 10^{-4}$ [27] |
| Na | 3.84 | Zr | 1.68 | Mg | 0.32 | Si | 2.2 | P | 0.8 | O | 12 | 3.84 | 0.430 | $5.06 \times 10^{-6}$ [27] |
| Na | 1 | Zr | 2 | | | | | P | 3 | O | 12 | 1 | 0.390 | $1.71 \times 10^{-8}$ [28] |
| Na | 1.5 | Zr | 1.5 | Al | 0.5 | | | P | 3 | O | 12 | 1.5 | 0.372 | $4.31 \times 10^{-8}$ [28] |
| Na | 1.5 | Zr | 1.5 | Cr | 0.5 | | | P | 3 | O | 12 | 1.5 | 0.380 | $2.17 \times 10^{-7}$ [28] |
| Na | 1.5 | Zr | 1.5 | Ga | 0.5 | | | P | 3 | O | 12 | 1.5 | 0.380 | $3.02 \times 10^{-8}$ [28] |
| Na | 1.5 | Zr | 1.5 | In | 0.5 | | | P | 3 | O | 12 | 1.5 | 0.398 | $3.93 \times 10^{-7}$ [28] |
| Na | 1.5 | Zr | 1.5 | Sc | 0.5 | | | P | 3 | O | 12 | 1.5 | 0.393 | $1.30 \times 10^{-6}$ [28] |
| Na | 1.5 | Zr | 1.5 | Y | 0.5 | | | P | 3 | O | 12 | 1.5 | 0.408 | $1.99 \times 10^{-6}$ [28] |
| Na | 1.5 | Zr | 1.5 | Yb | 0.5 | | | P | 3 | O | 12 | 1.5 | 0.405 | $6.67 \times 10^{-7}$ [28] |
| Na | 2 | Zr | 1 | Al | 1 | | | P | 3 | O | 12 | 2 | 0.353 | $5.57 \times 10^{-9}$ [28] |
| Na | 2 | Zr | 1 | Cr | 1 | | | P | 3 | O | 12 | 2 | 0.369 | $6.54 \times 10^{-7}$ [28] |
| Na | 2 | Zr | 1 | Ga | 1 | | | P | 3 | O | 12 | 2 | 0.370 | $2.85 \times 10^{-9}$ [28] |
| Na | 2 | Zr | 1 | In | 1 | | | P | 3 | O | 12 | 2 | 0.406 | $3.16 \times 10^{-6}$ [28] |
| Na | 2 | Zr | 1 | Sc | 1 | | | P | 3 | O | 12 | 2 | 0.395 | $5.35 \times 10^{-6}$ [28] |
| Na | 2 | Zr | 1 | Y | 1 | | | P | 3 | O | 12 | 2 | 0.426 | $2.30 \times 10^{-7}$ [28] |
| Na | 2 | Zr | 1 | Yb | 1 | | | P | 3 | O | 12 | 2 | 0.420 | $3.98 \times 10^{-6}$ [28] |
| Na | 2.5 | Zr | 0.5 | Cr | 1.5 | | | P | 3 | O | 12 | 2.5 | 0.359 | $4.00 \times 10^{-6}$ [28] |
| Na | 2.5 | Zr | 0.5 | In | 1.5 | | | P | 3 | O | 12 | 2.5 | 0.414 | $1.39 \times 10^{-6}$ [28] |
| Na | 2.5 | Zr | 0.5 | Sc | 1.5 | | | P | 3 | O | 12 | 2.5 | 0.398 | $2.32 \times 10^{-5}$ [28] |
| Na | 2.5 | Zr | 0.5 | Y | 1.5 | | | P | 3 | O | 12 | 2.5 | 0.444 | $7.52 \times 10^{-7}$ [28] |
| Na | 2.5 | Zr | 0.5 | Yb | 1.5 | | | P | 3 | O | 12 | 2.5 | 0.434 | $3.11 \times 10^{-6}$ [28] |
| Na | 3 | In | 2 | | | | | P | 3 | O | 12 | 3 | 0.422 | $1.42 \times 10^{-7}$ [28] |
| Na | 3 | Cr | 2 | | | | | P | 3 | O | 12 | 3 | 0.348 | $2.43 \times 10^{-8}$ [28] |

| | | | | | | | | | | | |
|---|---|---|---|---|---|---|---|---|---|---|---|
| Na | 3 | Fe | 2 | | | P | 3 | O | 12 | 3 | 0.360 | $1.03 \times 10^{-7}$ [28] |
| Na | 3 | Sc | 2 | | | P | 3 | O | 12 | 3 | 0.400 | $5.16 \times 10^{-5}$ [28] |
| Na | 1 | Ge | 2 | | | P | 3 | O | 12 | 1 | 0.314 | $1.54 \times 10^{-15}$ [29] |
| Na | 1 | Ge | 1.5 | Ti | 0.5 | P | 3 | O | 12 | 1 | 0.322 | $8.35 \times 10^{-13}$ [29] |
| Na | 1 | Ge | 1 | Ti | 1 | P | 3 | O | 12 | 1 | 0.329 | $1.05 \times 10^{-11}$ [29] |
| Na | 1 | Ge | 0.5 | Ti | 1.5 | P | 3 | O | 12 | 1 | 0.337 | $3.10 \times 10^{-10}$ [29] |
| Na | 1 | Ti | 2 | | | P | 3 | O | 12 | 1 | 0.344 | $5.29 \times 10^{-9}$ [29] |
| Na | 1 | Sn | 2 | | | P | 3 | O | 12 | 1 | 0.378 | $2.81 \times 10^{-11}$ [29] |
| Na | 1 | Sn | 1.5 | Ti | 0.5 | P | 3 | O | 12 | 1 | 0.370 | $2.12 \times 10^{-11}$ [29] |
| Na | 1 | Sn | 1 | Ti | 1 | P | 3 | O | 12 | 1 | 0.361 | $9.39 \times 10^{-11}$ [29] |
| Na | 1 | Sn | 0.5 | Ti | 1.5 | P | 3 | O | 12 | 1 | 0.353 | $5.69 \times 10^{-10}$ [29] |
| Na | 1 | Sn | 2 | | | P | 3 | O | 12 | 1 | 0.378 | $2.81 \times 10^{-11}$ [29] |
| Na | 1 | Zr | 0.5 | Sn | 1.5 | P | 3 | O | 12 | 1 | 0.381 | $6.00 \times 10^{-11}$ [29] |
| Na | 1 | Zr | 1 | Sn | 1 | P | 3 | O | 12 | 1 | 0.384 | $2.93 \times 10^{-10}$ [29] |
| Na | 1 | Zr | 1.5 | Sn | 0.5 | P | 3 | O | 12 | 1 | 0.387 | $1.15 \times 10^{-9}$ [29] |
| Na | 1 | Zr | 2 | | | P | 3 | O | 12 | 1 | 0.390 | $1.71 \times 10^{-8}$ [29] |
| Na | 2 | Hf | 2 | | | P | 3 | O | 12 | 2 | 0.386 | $4.72 \times 10^{-9}$ [29] |
| Na | 1.5 | Zr | 1.5 | Ga | 0.5 | P | 3 | O | 12 | 1.5 | 0.380 | $4.03 \times 10^{-8}$ [30] |
| Na | 1.5 | Zr | 1.5 | Cr | 0.5 | P | 3 | O | 12 | 1.5 | 0.380 | $5.33 \times 10^{-7}$ [30] |
| Na | 1 | Ti | 2 | | | P | 3 | O | 12 | 1 | 0.344 | $2.51 \times 10^{-14}$ [31] |
| Na | 1 | Zr | 2 | | | P | 3 | O | 12 | 1 | 0.390 | $1.06 \times 10^{-8}$ [31] |
| Na | 1 | Zr | 1 | Nb | 1 | P | 3 | O | 12 | 1 | 0.374 | $1.06 \times 10^{-8}$ [31] |
| Na | 1 | Ti | 1 | Nb | 1 | P | 3 | O | 12 | 1 | 0.351 | $1.25 \times 10^{-6}$ [31] |
| Na | 1 | Sc | 1 | Nb | 1 | P | 3 | O | 12 | 1 | 0.379 | $1.16 \times 10^{-11}$ [32] |
| Na | 1.5 | Sc | 1 | Nb | 1 | P | 3 | O | 12 | 1.5 | 0.379 | $5.42 \times 10^{-7}$ [32] |
| Na | 2 | Sc | 1 | Nb | 1 | P | 3 | O | 12 | 2 | 0.379 | $8.39 \times 10^{-7}$ [32] |
| Na | 2.5 | Sc | 1 | Nb | 1 | P | 3 | O | 12 | 2.5 | 0.379 | $1.62 \times 10^{-6}$ [32] |
| Na | 3 | Sc | 1 | Nb | 1 | P | 3 | O | 12 | 3 | 0.379 | $2.13 \times 10^{-6}$ [32] |
| Na | 1 | Zr | 2 | | | P | 3 | O | 12 | 1 | 0.390 | $4.46 \times 10^{-9}$ [33] |
| Na | 1.4 | Zr | 1.6 | In | 0.4 | P | 3 | O | 12 | 1.4 | 0.396 | $1.43 \times 10^{-7}$ [33] |
| Na | 1.6 | Zr | 1.4 | In | 0.6 | P | 3 | O | 12 | 1.6 | 0.400 | $6.17 \times 10^{-7}$ [33] |
| Na | 1.8 | Zr | 1.2 | In | 0.8 | P | 3 | O | 12 | 1.8 | 0.403 | $8.86 \times 10^{-7}$ [33] |
| Na | 2 | Zr | 1 | In | 1 | P | 3 | O | 12 | 2 | 0.406 | $2.20 \times 10^{-6}$ [33] |
| Na | 2.2 | Zr | 0.8 | In | 1.2 | P | 3 | O | 12 | 2.2 | 0.409 | $2.48 \times 10^{-6}$ [33] |
| Na | 2.4 | Zr | 0.6 | In | 1.4 | P | 3 | O | 12 | 2.4 | 0.412 | $3.77 \times 10^{-6}$ [33] |
| Na | 1.2 | Zr | 1.8 | Yb | 0.2 | P | 3 | O | 12 | 1.2 | 0.396 | $1.56 \times 10^{-7}$ [33] |
| Na | 1.4 | Zr | 1.6 | Yb | 0.4 | P | 3 | O | 12 | 1.4 | 0.402 | $2.19 \times 10^{-7}$ [33] |
| Na | 1.6 | Zr | 1.4 | Yb | 0.6 | P | 3 | O | 12 | 1.6 | 0.408 | $7.59 \times 10^{-7}$ [33] |
| Na | 1.8 | Zr | 1.2 | Yb | 0.8 | P | 3 | O | 12 | 1.8 | 0.414 | $2.68 \times 10^{-6}$ [33] |
| Na | 2 | Zr | 1 | Yb | 1 | P | 3 | O | 12 | 2 | 0.420 | $1.83 \times 10^{-6}$ [33] |
| Na | 2.4 | Zr | 0.6 | Yb | 1.4 | P | 3 | O | 12 | 2.4 | 0.431 | $1.95 \times 10^{-6}$ [33] |
| Na | 2.6 | Zr | 0.4 | Yb | 1.6 | P | 3 | O | 12 | 2.6 | 0.437 | $1.56 \times 10^{-6}$ [33] |
| Na | 2.8 | Zr | 0.2 | Yb | 1.8 | P | 3 | O | 12 | 2.8 | 0.443 | $3.66 \times 10^{-8}$ [33] |
| Na | 1.4 | Al | 0.4 | Ti | 1.6 | P | 3 | O | 12 | 1.4 | 0.338 | $1.06 \times 10^{-7}$ [34] |
| Na | 1.4 | Al | 0.4 | Sn | 1.6 | P | 3 | O | 12 | 1.4 | 0.366 | $1.24 \times 10^{-8}$ [34] |
| Na | 1.4 | Al | 0.4 | Ge | 1.6 | P | 3 | O | 12 | 1.4 | 0.314 | $6.64 \times 10^{-10}$ [34] |